\documentclass[prd,twocolumn,showpacs,preprintnumbers,amsmath,amssymb]{revtex4}
\usepackage{graphicx}
\usepackage{epsfig}
\usepackage{rotating}
\usepackage{dcolumn}
\usepackage{bm}

\newcommand{\gsim}{\;\raisebox{-0.9ex}
           {$\textstyle\stackrel{\textstyle >}{\sim}$}\;}
\newcommand{\lsim}{\;\raisebox{-0.9ex}{$\textstyle\stackrel{\textstyle<}
           {\sim}$}\;}

\begin{document}

\preprint{hep-ph/0609113}
\preprint{IISc-CHEP/4/06}
\preprint{CERN-PH-TH/2006-117}
\preprint{LPT-ORSAY-06-56}

\title{Probing CP-violating Higgs contributions in $\gamma\gamma\to f\bar f$ through fermion polarization}
\author{Rohini M. Godbole}
\email{rohini@cts.iisc.ernet.in}
\affiliation{Center for High Energy Physics, IISc, Bangalore 560 012, India}
\author{Sabine Kraml}
\email{Sabine.Kraml@cern.ch}
\affiliation{Theory Division, Department of Physics, CERN, CH-1211 Geneva 23, Switzerland}
\author{Saurabh D. Rindani}
\email{saurabh@prl.res.in}
\affiliation{Physical Research Laboratory, Navrangapura, Ahmedabad 380 009, India}
\author{Ritesh K. Singh}
\email{Ritesh.Singh@th.u-psud.fr}
\affiliation{Laboratoire de Physique Th\'eorique, 91405 Orsay Cedex, France}

\begin{abstract}
We discuss the use of fermion polarization for studying neutral Higgs bosons at
a photon collider. 
To this aim we construct polarization asymmetries which can isolate the 
contribution of a Higgs boson $\phi$ in $\gamma\gamma\to f \bar f$, $f=\tau/t$,
from that of the QED continuum. 
This can help in getting information on the $\gamma\gamma\phi$ coupling 
in case $\phi$ is a CP eigenstate. We also construct CP-violating 
asymmetries which can probe CP mixing in case $\phi$ has indeterminate CP. 
Furthermore, we take the MSSM with CP violation as an example to demonstrate 
the potential of these asymmetries in a numerical analysis.  
We find that these asymmetries are sensitive to the presence of a Higgs
boson as well as its CP properties over a wide range of MSSM parameters. 
In particular, the method suggested can cover the region where a light 
Higgs boson may have been missed by LEP due to CP violation in the 
Higgs sector, and may be missed as well at the LHC.
\end{abstract}
\maketitle

\section{Introduction}

The Higgs boson is the only particle  
of the Standard Model (SM) to have eluded experimental discovery so far. 
The discovery of the Higgs boson
and the subsequent study of electroweak symmetry breaking is one of 
the prime aims of all the current and next generation colliders 
\cite{Godbole:2002mt}. 
Electroweak precision measurements indicate, in the SM, the existence of a Higgs
boson lighter than 204 GeV at 95\%~C.L.~\cite{higmax}. A Higgs boson with SM
couplings lighter than 114.4 GeV is ruled out by direct searches at
LEP~\cite{higmin}. Thus one expects to find the SM Higgs boson with a mass in
this range. In models with an extended (and possibly CP-violating) Higgs sector,
the couplings of electroweak vector bosons to the lightest Higgs can be
suppressed~\cite{gvvh1}. In such a case direct searches allow the existence of 
a Higgs boson much lighter than 114.4 GeV~\cite{Abbiendi:2004ww}.

The Large Hadron Collider (LHC), scheduled to go in operation in 2007, is
expected~\cite{tdr} to be capable of searching for the SM Higgs boson in the
entire mass range expected theoretically and still allowed experimentally. 
The International Linear Collider (ILC)~\cite{ilc}, currently in planning,  
is expected to be capable of profiling the Higgs boson very accurately, 
again for the entire mass range mentioned above. 
Determination of the CP properties of the spin-0 particle, which
we hope will be discovered and studied at the LHC and the ILC, is an important
part of this project of profiling the Higgs boson, see e.g. \cite{cpnsh}. 
The Higgs couplings 
with a pair of electroweak gauge bosons $(V=W/Z)$ and those with a pair 
of heavy fermions $(f = t/\tau)$ are the ones that prove the most useful in 
this context. These couplings, for a neutral Higgs boson $\phi$, which may 
or may not be a CP eigenstate, can be written as
\begin{eqnarray}
  \phi f\bar{f} &:& \frac{-i g \; m_f}{2 \ M_W} \ 
   \left( v_f + i a_f\,\gamma_5 \right)\label{v1}\\[2mm]
  \phi VV &:& \frac{i g \; M_V^2}{M_W} \ 
              \left(A_V g_{\mu\nu}  
                    + B_V \frac{p_\mu p_\nu}{M_Z^2} 
                    + i\,C_V \epsilon_{\mu\nu\rho\sigma} 
                      \frac{p^\rho q^\sigma}{M_Z^2} \right)\nonumber\\
  \label{ffVV}
\end{eqnarray}
where $p = P_{V_1} + P_{V_2}$,  $q = P_{V_1} - P_{V_2}$ and $P_{V_1}, \
P_{V_2}$ are the four momenta of the two massive vector bosons. 
In the SM, $v_f = A_V =1$ and $a_f = B_V = C_V = 0$. 
At the LHC, the $t\bar t$ final state produced in the decay of an inclusively 
produced Higgs can provide knowledge of the CP nature of the $t\bar t\phi$ 
coupling 
through spin-spin correlations~\cite{spin2} whereas $t\bar t\phi$ production 
can allow a determination of $v_f$ and $a_f$~\cite{s2LHC}. 
It should also be possible to exploit the $\phi ZZ$ coupling 
via $\phi \to ZZ \to l^+l^- l'^+l'^-$ \cite{zz}, 
and the vector boson fusion mode~\cite{plehn}. 
At the ILC~\cite{ilc}, a rather clean determination of the CP of the 
Higgs boson should be possible using the Higgsstrahlung process~\cite{hzzLC}. 
Angular correlations of the decay products of the $\phi$, 
in particular $\phi\to f \bar f$, may be used effectively 
at an $e^+e^-$ collider 
to distinguish between $v_f=1,a_f=0$ and $v_f=0,a_f=1$ as well as to get 
information on CP mixing~\cite{s2LC}.
A remark is in order here. The different methods which exploit the $VV\phi$
coupling test the terms with tensor structure $g_{\mu\nu}$ and 
$\epsilon_{\mu\nu\rho\sigma}$ in Eq.~(\ref{ffVV}); actually in most cases 
the CP-even part $\propto g_{\mu\nu}$ is projected out since in most 
CP violating (CPV) extensions of the SM one has $A_V \gg B_V, C_V$. 
The $f\bar f\phi$ coupling, on the 
other hand, allows equal sensitivity to the CP-even and CP-odd parts. At 
both colliders, the LHC and the ILC, the determination of the CP quantum 
number of the Higgs  boson  seems feasible, while determination of CP mixing 
seems difficult; 
the best chance for the latter being offered by the
exploitation of the $f\bar f\phi$ coupling~\cite{Godbole:2004xe}.

A more sensitive laboratory for studying the CP properties of a neutral 
Higgs boson is the photon collider option \cite{plc} of the ILC.  
At a photon collider, the Higgs
boson is produced resonantly in the $s$-channel and thus one will be able to 
produce it copiously even if the $\gamma\gamma \phi$ coupling is small.  In 
presence of CP violation (CPV), the CP-even and CP-odd components of the Higgs couple 
to photons with comparable strength. A study of the production rates with 
linearly- and circularly-polarized photons can help determine the CP quantum 
numbers if conserved, as well as the CP mixing phases in the case of CP 
violation. Angular distributions of decay products ($VV, \ b\bar b$) of the 
produced $\phi$ along with measurements of $\Gamma(\phi\to b\bar b), \ \Gamma(\phi\to\gamma\gamma)$ 
may also allow the determination of a CP-mixing phase~\cite{maria}; 
the $\phi\to t\bar t$ decay can also be used~\cite{hel-Asa,hel-God} 
if it is kinematically allowed.
While a complete reconstruction of couplings would require both 
circularly and linearly polarized photons~\cite{hel-Asa}, substantial 
information can already be obtained using circularly polarized photons 
alone~\cite{hel-God}.
\begin{figure}
\begin{center}
\includegraphics[scale=0.60]{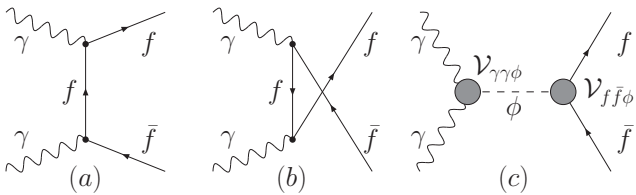}
\caption{\label{feynman}
Feynman diagrams contributing to $\gamma \gamma \rightarrow f \bar f$
production.}\end{center}
\end{figure}
\noindent
Polarization of the final state fermions, which are pair produced, can be a 
probe of CP violation in the production process~\cite{ma,hel-Asa,hel-God}.  
Among SM fermions, the polarization can be measured only for t-quarks and
tau-leptons. The $t$-quark decays before 
it hadronizes and  hence the polarization information gets reflected in the 
angular distributions of decay products, whereas for the $\tau$ lepton one can 
extract polarization information by looking at the energy distribution of 
pions~\cite{nojiri,Bullock:1992yt}. For a Higgs mass larger than $2 m_t$, top
polarization is the best probe of the Higgs interaction. However, for a 
CP-violating light Higgs boson, which could have escaped detection at LEP,
the decay into  $t\bar t$  is not possible. In this case the $\tau^+ \tau^-$
final state might provide a useful probe.  
We therefore study the contribution of a light Higgs boson 
to $\tau$ production via $\gamma\gamma\to\phi\to\tau^+\tau^-$ and that of
heavier Higgs bosons to $t$ production via $\gamma\gamma\to\phi\to t\bar t$.
Higgs contribution to $\tau$ polarization has also been studied recently 
for the LHC \cite{coupled}, for a photon collider \cite{Ellis:2004hw} and for 
an $e^+e^-$ linear collider \cite{Ellis:2005ik} in the context of resonant CP
violation in the MSSM. The effect of neutral Higgs boson exchange in
$\gamma\gamma\to t\bar t$ has been addressed in 
Ref.~\cite{hel-Asa,hel-God,Choi:2004kq},
in terms of various correlations and asymmetries for a photon collider.

In this paper, we investigate the possibility of using tau/top polarization 
to get information on the $\phi f\bar f$ and $\phi\gamma\gamma$ couplings,  
as well as to probe possible CP violation in them in a generic scenario. 
First, we formulate the method in a model-independent way.  
We then apply it in a numerical analysis to the minimal supersymmetric standard 
model with CP violation (CPV-MSSM). 
The neutral Higgs sector of the CPV-MSSM has been studied in detail 
theoretically~\cite{cpviolHiggs}, and constraints from LEP are 
available~\cite{Abbiendi:2004ww}.  In this work, we study $\tau/t$ polarization 
in the CPX scenario~\cite{cpx} over regions of the CPV-MSSM parameter space 
that are allowed by the current data, and assess the feasibility of using it  
to probe the Higgs contribution to $f\bar f$ production. 
In the numerical analysis, we use {\tt CPsuperH}~\cite{CPsuperH} and 
{\tt FeynHiggs 2.1}~\cite{FeynHiggs}
for calculating the masses, the decay 
widths and the relevant couplings of the  Higgs bosons.

The rest of the paper is organized as follows. In section \ref{two}, we
describe fermion pair production at a $\gamma\gamma$ collider in a 
model-independent way. Section \ref{three} deals with polarization 
observables as potential 
probes of the Higgs contribution and its CP structure. Numerical results 
within the CPV-MSSM are presented in section \ref{four}, and in section 
\ref{five} we present our conclusions.

\section{Fermion pair production in $\gamma\gamma$ collision}
\label{two}

At a photon collider, the production of a fermion pair involves the 
$\gamma f\bar f$, $\gamma\gamma\phi$ and $\phi f\bar f$ vertices
as shown in Fig.~\ref{feynman}. 
We take the $\gamma f\bar f$ vertex to be the standard QED one, while the 
vertices involving Higgs bosons are taken to be the most general allowed. 
The model independent vertex for Higgs interactions with fermions is given in
Eq.~(\ref{v1}) and with a pair of photons, allowing for CP violation, 
can be written as~:
\begin{eqnarray}
   {\cal V}_{\gamma\gamma\phi}^{\mu\nu}
   &=& \frac{-i\sqrt{s}\alpha}{4\pi}\left[A_{\gamma}(s)
       \left(g^{\mu\nu}-\frac{2}{s}k_1^\nu \ k_2^\mu 
       \right) \right.\nonumber\\
   & & \hspace{14mm}-\,\left. B_{\gamma}(s)\frac{2}{s}
       \epsilon^{\mu\nu\alpha\beta}
       k_{1\alpha} k_{2\beta} \right].
\label{v2}
\end{eqnarray}
Here $k_1$ and $k_2$ are the four-momenta of the colliding photons. The
helicity amplitudes for fermion pair production in the $s$- and the 
$t/u$-channels can be obtained from those given in \cite{hel-Asa,hel-God}~:
\begin{eqnarray}
  && M_{\phi}(\lambda_1,\lambda_2;\lambda_f,\lambda_{\bar f}) 
     = \frac{-ig\alpha m_f}{8\pi M_W} \ 
       \frac{s}{s-m_{\phi}^2+im_{\phi}\Gamma_{\phi}} \nonumber\\
  && \left[ A_{\gamma}(s) + i\lambda_1 B_{\gamma}(s) \right]
     \left[ \lambda_f \beta v_f - ia_f \right] \ 
     \delta_{\lambda_1,\lambda_2} \delta_{\lambda_f,\lambda_{\bar f}} \ ,
     \label{prodhig}
\end{eqnarray}
\begin{eqnarray}
  && M_{QED}(\lambda_1,\lambda_2;\lambda_f,\lambda_{\bar f})
     = \frac{-i4\pi\alpha Q^2}{1-\beta^2 \cos^2\theta_f} \nonumber\\
  && \left[\frac{4m_f}{\sqrt{s}} \ (\lambda_1+\lambda_f\beta)
           \ \delta_{\lambda_1,\lambda_2}\delta_{\lambda_f,\lambda_{\bar f}}
     \right.\nonumber\\
  && \;\; -\frac{4m_f}{\sqrt{s}} \ \lambda_f \beta\sin^2\theta_f \
    \delta_{\lambda_1,-\lambda_2}\delta_{\lambda_f,\lambda_{\bar f}}\nonumber\\
  && \;\; - 2\beta \ (\cos\theta_f + \lambda_1\lambda_f)\sin\theta_f \ 
     \delta_{\lambda_1,-\lambda_2}\delta_{\lambda_f,-\lambda_{\bar f}}
     \Big].
\label{prodsm}
\end{eqnarray}
The form factors $A_\gamma$ and $B_\gamma$ are complex whereas $v_f, a_f$ 
can be taken to be real without loss of generality. The non-standard 
vertices given by Eqs.~(\ref{v1}) and (\ref{v2}) involve four independent 
form factors: $v_f,\ a_f,\ A_{\gamma},\ B_{\gamma}$. 
In the MSSM with CPV, these form factors are functions of various model 
parameters: $\tan\beta$; $m_{H^+}$; $(|\mu|,\Phi_\mu)$; 
$(|A_{\tilde{f}}|,\Phi_{\tilde{f}})$; $(|M_i|,\Phi_i)$, $i=1,2,3$;  
$m_{\tilde{q},\tilde{l}}$; etc., where $(|x|,\Phi_x)$ denotes  
$x=|x|e^{i\Phi_x}$.

The helicity amplitudes of Eqs.~(\ref{prodhig}) and (\ref{prodsm}) 
involve only certain combinations 
of the form factors which are listed in Table~\ref{comb}.
%
\begin{table}
\caption{\label{comb}Combinations of form factors $v_f, \ a_f,
\ A_{\gamma}$ and $B_{\gamma}$ that occur in the helicity amplitudes of  
Eqs.~(\ref{prodhig}) and (\ref{prodsm}).}
\begin{ruledtabular}
\begin{tabular}{ccc|ccc}
Combination & Alias & CP& Combination & Alias & CP\\
\hline
$v_f \Re(A_\gamma)$ & $x_1$ & even &
$v_f \Im(A_\gamma)$ & $x_2$ & even \\
$v_f \Re(B_\gamma)$ & $y_1$ & odd  &
$v_f \Im(B_\gamma)$ & $y_2$ & odd  \\
$a_f \Re(A_\gamma)$ & $y_3$ & odd  &
$a_f \Im(A_\gamma)$ & $y_4$ & odd  \\
$a_f \Re(B_\gamma)$ & $x_3$ & even &
$a_f \Im(B_\gamma)$ & $x_4$ & even \\ 
\end{tabular}
\end{ruledtabular}
\end{table}
%
Only five of these eight combinations are independent, the other three can be 
obtained by inter-relations such as $x_1x_3 = y_1y_3$, etc. In all the 
extensions of the SM, $A_\gamma$ and $B_\gamma$ are generated at the one-loop
level. Simultaneous existence of $v_f$ and $a_f$, or $A_\gamma$ and $B_\gamma$
violates CP, i.e.\ non-vanishing values of $y_i$, $(i=1,...,4)$ imply CP
violation. Even in case of CP invariance, where only the $x_i$'s are 
non-zero, the Higgs contribution can alter the polarization of the fermions 
$f$ from that predicted by pure QED. CP violation, giving
rise to non-zero $y_i$'s, gives an additional contribution to the fermion 
polarization.

It should be noted that the Higgs-mediated diagram contributes only when the 
helicities of the colliding photons are equal. The helicities of $f$ and
$\bar f$ are also equal in this case. The QED contribution for this helicity 
combination is proportional to the fermion mass. Both these facts indicate 
that one should choose equal photon helicities to enhance the effect of the 
Higgs mediated diagram. The contribution with opposite helicities of photons 
comes from QED diagrams alone and it is large as compared to that of equal 
photon helicities for $\sqrt{s}\gg 4m_f$. 
Thus with unpolarized photons the net 
contribution from Higgs exchange will be relatively small. Hence 
one expects poor sensitivity to the Higgs contribution with unpolarized 
initial state photons.

Among the vertices contributing to fermion-pair production, the standard 
$\gamma f\bar f$ vertex conserves chirality, while the $\phi f \bar f$ vertex 
mixes different chiralities. Owing to the finite mass $m_f$ of the fermion, there 
is a  chirality-mixing contribution even for the pure QED diagrams 
(Figs.~\ref{feynman}a,b). The presence of the Higgs boson 
exchange, Fig.~\ref{feynman}c, provides an additional, polarization dependent, 
spatially isotropic, chirality-mixing contribution. The property of spatial 
isotropy is unique to Higgs exchange contribution, in contrast to other means 
of chirality mixing, such as the finite  mass effect. 
With unpolarized initial-state photons,
the QED as well as a CP-conserving Higgs contribution lead to
unpolarized fermions in the final state; CP violation in the Higgs sector
leads to a net, though very small, fermion polarization.
With polarized initial-state photons, already pure QED leads to a finite
polarization. The additional chirality mixing from the Higgs exchange
causes a change in this polarization in both the CP-conserving and the
CP-violating case. It is thus possible  to construct observables relating
initial-state photon and final-state $\tau/t$ polarizations which probe
the Higgs couplings as well as possible CP violation in the Higgs sector.

\section{Fermion polarization in $\gamma\gamma$ collision}
\label{three}

\begin{figure}
\epsfig{file=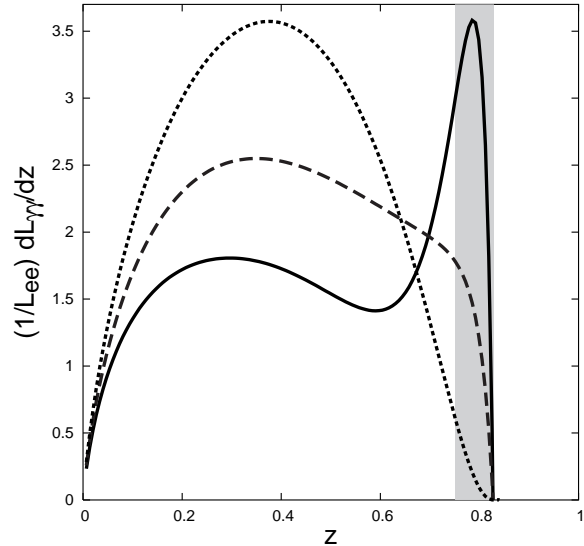,width=8.0cm}
\caption{\label{spec}
Luminosity distribution plotted against $z$ (which is related to the
$\gamma\gamma$ invariant mass $W=2\sqrt{\omega_1\omega_2}$ via $z=W/(2E_b)$)
for $x_c=4.8$. The solid line corresponds to $\lambda_e\lambda_l=-1$, 
the short-dashed line is for $\lambda_e\lambda_l=1$, and the long-dashed one 
for $\lambda_e\lambda_l=0$. The conversion distance is taken to be zero. 
The grey patch highlights the region $0.75<z<0.83$.}
\end{figure}

At a $\gamma\gamma$ collider, Compton back-scattering of a laser from 
$e^-/e^+$ is used to produce high-energy photons\cite{ginzburg}. 
The energy spectrum of 
the back-scattered photons depends upon the polarizations of the $e^-/e^+$ 
beams and the laser as shown in Fig.~\ref{spec}. 
This figure shows the distribution of the reduced invariant mass of the 
$\gamma\gamma$ system 
$z = \sqrt{\omega_1\omega_2}/E_b$, where $\omega_{1,2}$
are the energies of the two colliding photons in the lab frame, for a laser
energy $\omega_0$ corresponding to  
\begin{equation}
x_c = \frac{4 E_b \omega_{0}}{m_e^2} = 4.8.
\end{equation}
The spectrum is peaked in the high
energy region for opposite polarizations of the electron  and laser beams. 
Further, most of
these high-energy photons have a large degree of polarization. 
By choosing appropriate polarizations for the $e^-/e^+$ and laser beams, 
one can thus obtain a peaked and highly polarized spectrum for the 
colliding photons. We shall use this
fact when constructing various observables with $\tau$ polarization.

The polarization of fermions is defined as the fractional surplus of positive
helicity fermions over negative helicity ones, i.e.
\begin{equation}
   P_{f}^{ij} = \frac{N^{ij}_+ - N^{ij}_-}{N^{ij}_+ + N^{ij}_-} \ ,
\end{equation}
where the superscript $ij$ stands for the polarizations of the parent 
$e^-, \ e^+$ beams of the ILC (with $P_f^{++}$ meaning 100\%
right polarized electrons and 100\% right polarized positrons);
$N_+$ and $N_-$ stands for the number of fermions 
with positive and negative helicities, respectively. 
Analogously, $\bar{P}_f^{ij}$ is the polarization of the anti-fermion.
Due to the left-chiral 
nature of the $W$--boson interaction, the positively and negatively polarized 
$f$'s lead to different distributions of the decay products. For $\tau$'s, by
looking at the net energy distribution of decay $\pi$ one can get information
on its polarization~\cite{nojiri}. On the other hand, for $t$-quarks, it is the
energy distribution of $b$-quarks or the angular distribution of decay leptons.

The various $N^{ij}_\pm$, and hence $P_f^{ij}$, with different polarizations 
of initial-state photons are related to each other via discrete symmetry 
transformations, such as C, P and CP, if these symmetries are respected by the 
underlying dynamics. Thus any deviation from these relations can be a probe of 
the violation of the corresponding discrete symmetry. 

For unpolarized initial-state photons the polarization, 
$P_f^U$, is zero for the QED diagrams (Figs.~\ref{feynman}a,b). 
Even in the presence of a CP-conserving Higgs $P_f^U$ is zero. This is 
because the left chiral and the right chiral components of fermions couple to 
the Higgs boson with equal strength. Thus the net polarization of the fermions, 
if any, will signal CP violation in the Higgs sector. In general it is a 
signal of P violation, but in the process under consideration $f$ couples 
only to self-conjugate neutral particles, $\gamma$ and $\phi$. 
Hence $P_f^U\neq0$ is also a signal of CP violation. 
We note that $P_f^U$ is a pure but a poor 
probe of CPV.

For polarized initial state photons, QED predicts a non-zero value of $P_f$ 
and this prediction is modified by the presence of the Higgs exchange
diagram. The deviation from the QED prediction is a probe of the Higgs 
contribution and hence its couplings. We define
\begin{eqnarray}
\delta P_f^+  &=& P^{++}_f - (P^{++}_f)^{QED}, 
\label{obs1}
\\
\delta P_f^-  &=& P^{--}_f - (P^{--}_f)^{QED}.
\label{obs2}
\end{eqnarray}
Such a deviation does not have any 
definite P or CP property, thus it can be non-zero even when the Higgs boson 
is a CP eigenstate. This allows us to detect the presence of a Higgs over 
a large range of $x_i$'s and $y_i$'s (and hence a large range of model 
parameters, as we will see later)
by measuring the fermion $(t /\tau)$ polarization.
We choose equal polarization of $e^+$ and $e^-$ beams so as to have equal
helicities for the colliding photons; this enhances the chirality-mixing 
Higgs contribution as discussed in the previous 
section, c.f.\ Eqs.~(\ref{prodhig}) and (\ref{prodsm}). 

In QED, the polarization of $f$ flips its sign if we change 
the signs of the polarizations of the initial state photons 
(actually those of the electron and positron of the parent collider), 
i.e. $P^{++}_f = -P^{--}_f$. 
This is due to P invariance of QED. Also, due to the self-conjugate 
nature of the neutral particles involved, 
we have $P^{ii}_f = \bar{P}^{ii}_f$. 
In CP-violating models, however, we expect $P^{++}_f + P^{--}_f \neq 0$. 
Although it is a probe of P violation in general, it will be a probe of 
CP violation in our process. In Table~\ref{obs} 
we list all the observables and their potentials.
%
\begin{table}
\caption{\label{obs}Polarization observables and interactions 
and combinations that they can probe.}
\begin{ruledtabular}
\begin{tabular}{c|c|c}
Observables & Interactions & Combinations\\ 
 & probed & probed \\ \hline
$P_f^{U}$ &
$P/CP$ violating & $y_i$'s\\
$\delta P_f^+ = P^{++}_f - (P^{++}_f)^{QED}$ &
Chirality-mixing &$x_i$'s, $y_i$'s \\
$\delta P_f^- = P^{--}_f - (P^{--}_f)^{QED}$ &
Chirality-mixing &$x_i$'s, $y_i$'s \\
$\delta P_f^{\rm CP} = P^{++}_f + P^{--}_f$ &
$P/CP$ violating & $y_i$'s\\ 
\end{tabular}
\end{ruledtabular}
\end{table}
\begin{figure}
\epsfig{file=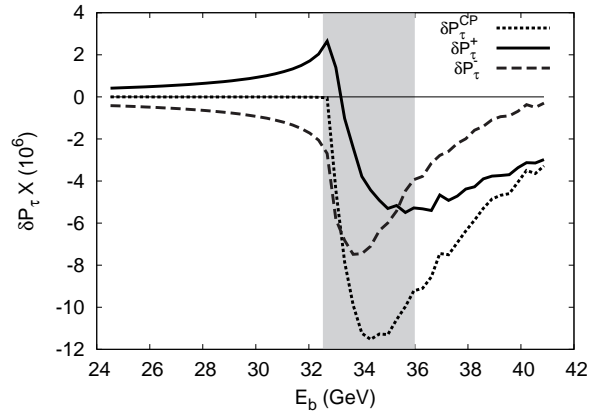,width=8.0cm}
\caption{\label{p-eb}
Variation of $\delta P^+$, $\delta P^-$ and $\delta P^{\rm CP}$ as a function
of $E_b$ for a Higgs boson of mass  54 GeV. For a value of $E_b$ in the 
grey patch, the mass of the Higgs boson matches with the 
$\gamma\gamma$ invariant  mass in the grey band shown in Fig.~\ref{spec}.}
\end{figure}
\begin{figure}[!htb]
\begin{tabular}{ccc}
{\tt CPsuperH} && {\tt FeynHiggs}\\
\epsfig{file=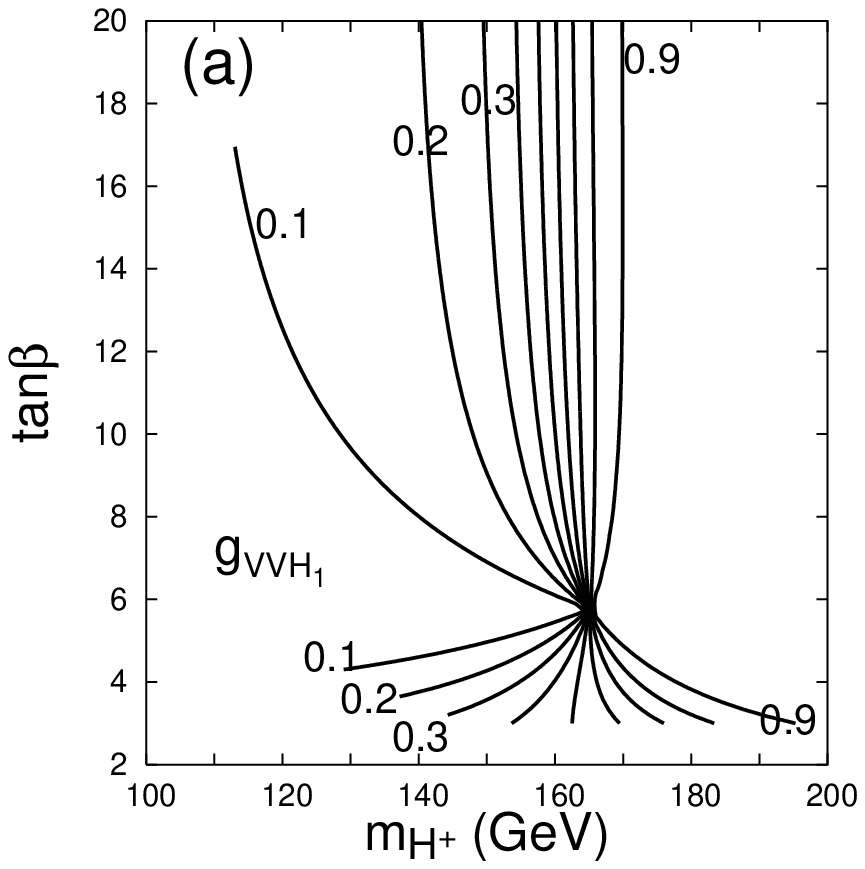,width=4.0cm,height=3.8cm}&  
~ &
\epsfig{file=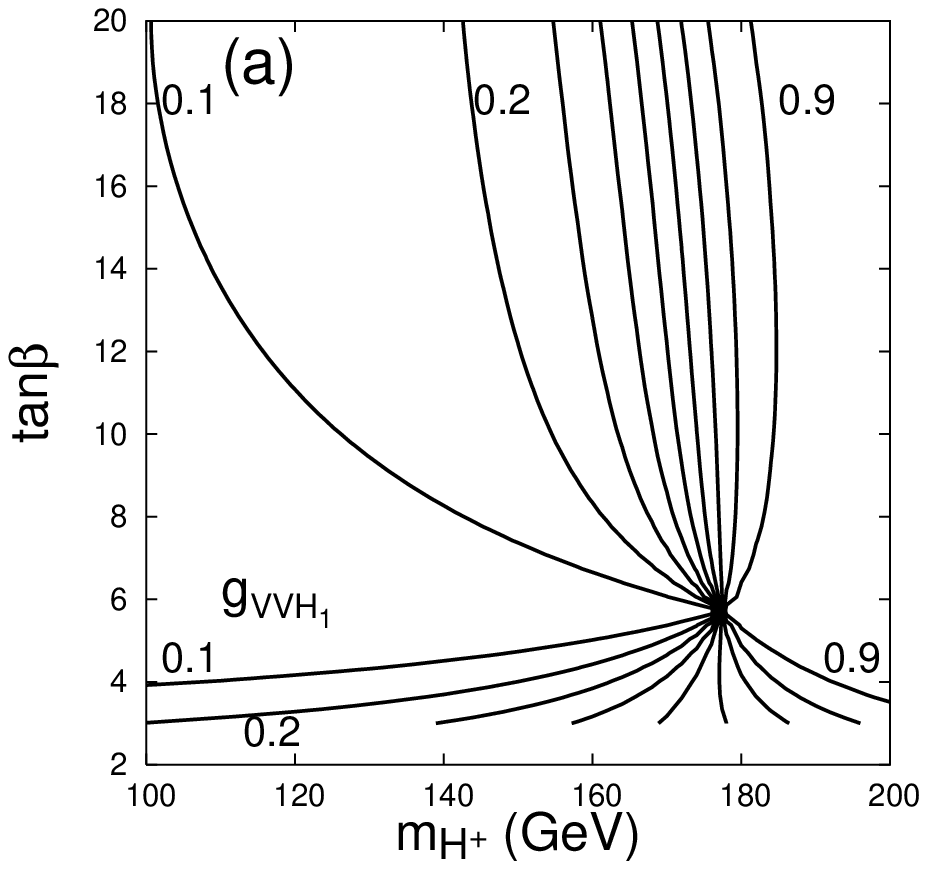,width=4.0cm,height=3.8cm} \\
\epsfig{file=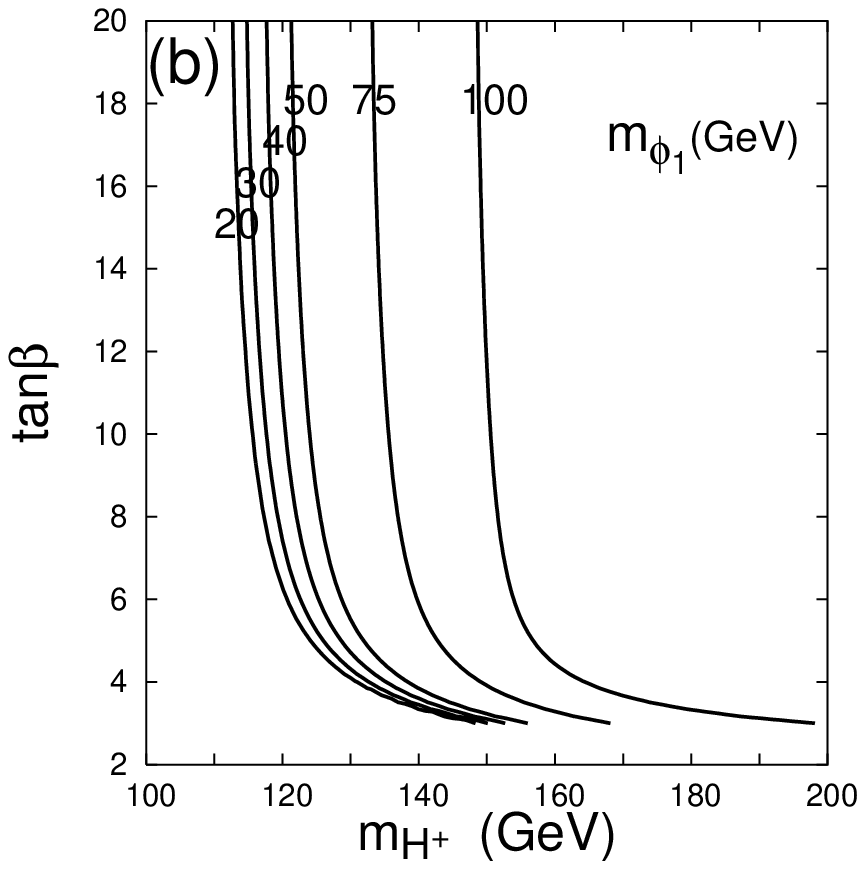,width=4.0cm,height=3.8cm}&  
~ &
\epsfig{file=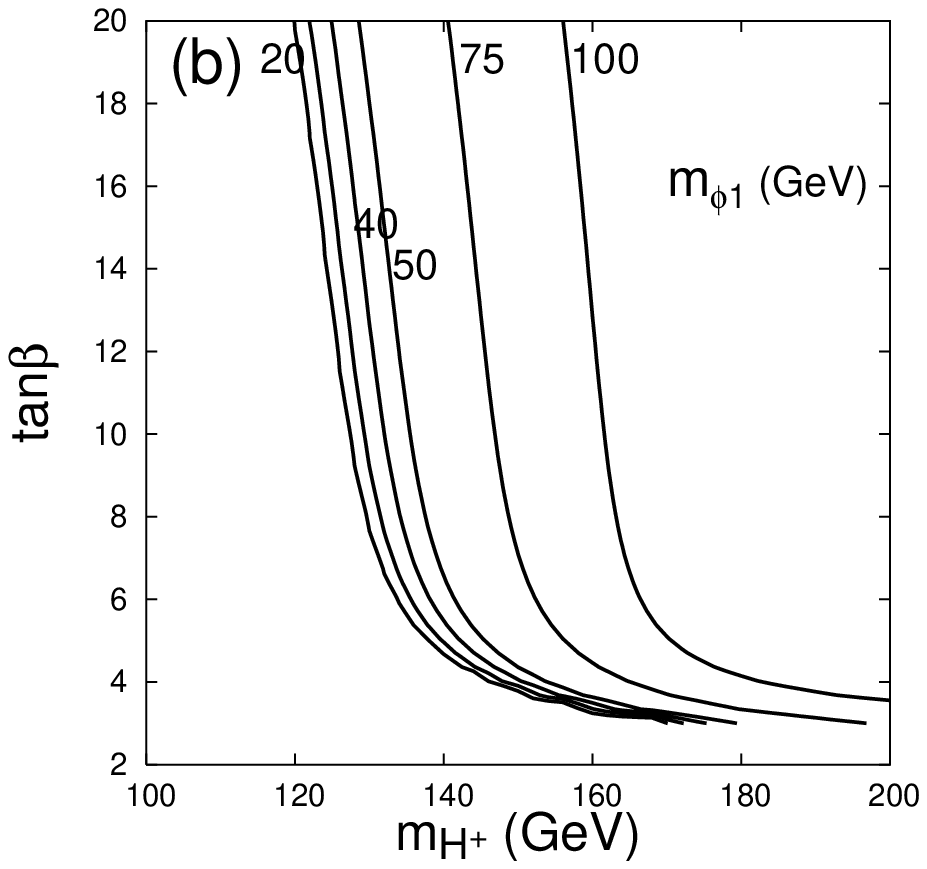,width=4.0cm,height=3.8cm} \\
\epsfig{file=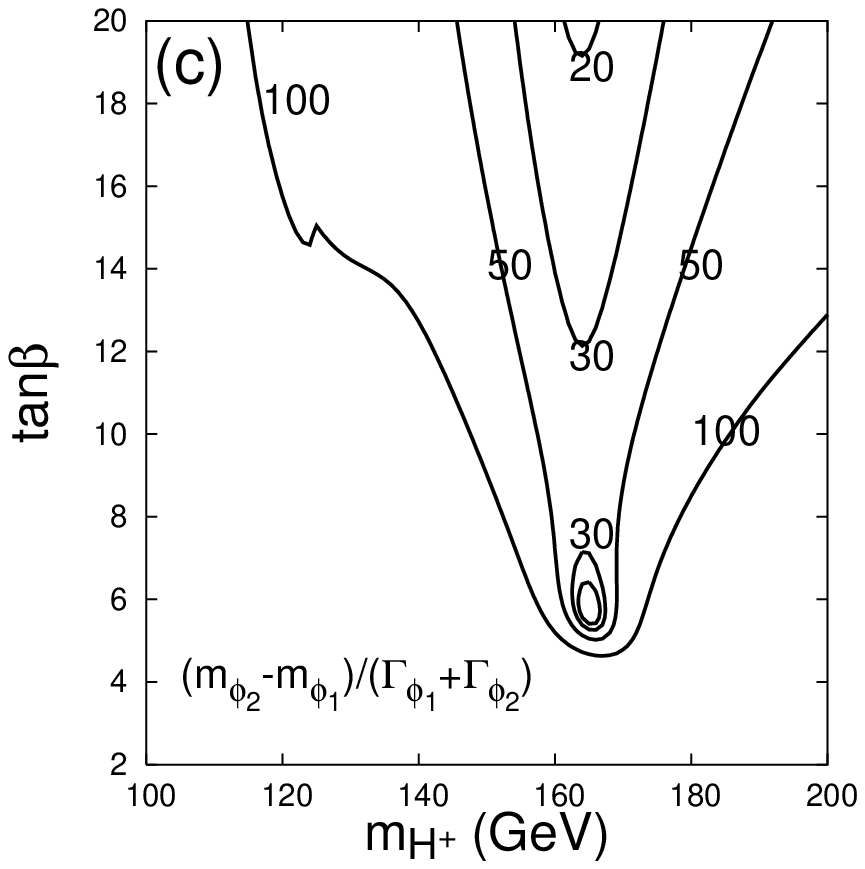,width=4.0cm,height=3.8cm}&	
~ &
\epsfig{file=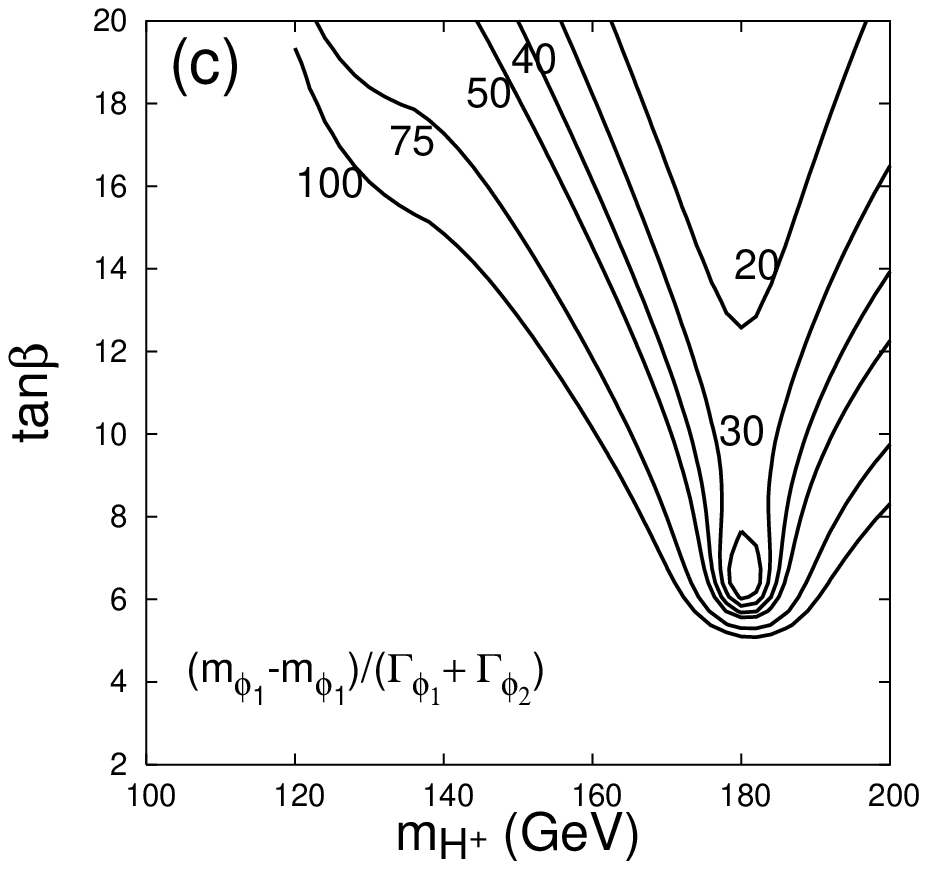,width=4.0cm,height=3.8cm}	
\end{tabular}
\caption{\label{cpxfig}CPX scenario: contours of (a) $A_V$ 
(see Eq.(2)) for $V=W,Z$, (b) $m_{\phi_1}$ and (c) 
$(m_{\phi_2}-m_{\phi_1})/(\Gamma_{\phi_1}+\Gamma_{\phi_2})$ 
computed with {\tt CPsuperH} as well as {\tt FeynHiggs}.}
\end{figure}
%
The definitions in Eqs.~(\ref{obs1}) and (\ref{obs2}), along with the fact
that $(P^{++}_f)^{QED} = -(P^{--}_f)^{QED}$, implies
\begin{equation} 
   \delta P_f^{\rm CP} = \delta P_f^+ + \delta P_f^- .
   \label{sum-rule}
\end{equation} 
For polarized photons we therefore have two independent observables, 
$\delta P_f^-$ and $ \delta P_f^{\rm CP}$. These can be sizable 
over a large range of, for instance, MSSM parameters and hence can 
be used to probe the Higgs interactions. 
As we have already mentioned, the expected polarization $P_f^U$ 
for unpolarized photons is very small. In Fig.~\ref{p-eb} we show, 
as an example, expected values of $\delta P^\pm_\tau$ and $\delta 
P^{\rm CP}_\tau$ 
as functions of $E_b$ for a Higgs mass of 54 GeV and $v_\tau = 2.0$, 
$a_\tau=-2.3$, $A_\gamma = -0.76 + 0.032 i$, $B_\gamma = -0.13 + 0.039 i$. 
The peak in $\delta P$ 
occurs when the Higgs mass matches with the $\gamma\gamma$ invariant 
mass corresponding to the peak of photon spectrum in Fig.~\ref{spec}. 
By adjusting the beam energy $E_b$, we can thus maximize the sensitivity 
of the polarization observables defined above. Details of the selection 
criteria for $E_b$ for a scan over the parameters of a model  
are discussed in Section~\ref{five}.

\section{The CPV-MSSM Higgs sector}
\label{four}

We choose the MSSM as an example for demonstrating the potential 
of the observables constructed in the previous section to isolate the Higgs 
boson contribution and to probe the CP properties of its couplings.
In the CP-conserving MSSM, there exist three neutral Higgs bosons: the CP-even 
$h,H$ and the CP-odd $A$. CP-violating phases of the MSSM parameters such as
the higgsino mass parameter $\mu$, gaugino masses $M_{i}$ $(i=1,2,3)$ 
and trilinear couplings $A_f$ $(f=t,b,\tau)$, can induce CP violation 
in the Higgs sector via loops. This allows Higgs states with different 
CP to mix; the three mass eigenstates hence do not have definite CP. 
These states are denoted by $\phi_1$, $\phi_2$ and $\phi_3$ with their 
masses in increasing order. 
For the numerical analysis, we choose the so-called CPX~\cite{cpx}
scenario with parameters as listed in Table~\ref{mspara}. 
In this scenario, one can have large CP-violating effects in the  
Higgs sector depending upon the size of the phases of the trilinear 
couplings $A_{t,b,\tau}$.
%
\begin{table}
\caption{\label{mspara}List of MSSM parameters for the CPX scenario 
used as input for the programs {\tt CPsuperH} and {\tt FeynHiggs}.}
\begin{ruledtabular}
\begin{tabular}{c|l|c|l}
MSSM & Value & MSSM & Value\\ 
param. & & param. & \\\hline
$\tan\beta$     & 3\,--\,40 (for scan)&
$m_{H^+}$       & 150\,--\,500 GeV (for scan)\\
$\mu$           & 2 TeV, $\Phi_\mu$ = 0&
$M_1,M_2$       & 200 GeV, $\Phi_{1,2}$ = 0\\
$M_3$           & 1 TeV, $\Phi_{3}$ = 90$^\circ$&
$m_{\tilde q,\tilde l}$ & 500 GeV\\
$A_{t,b}$       & 1 TeV, $\Phi_{t,b}$ = 90$^\circ$&
$A_\tau$        & 500 GeV, $\Phi_{\tau}$ = 90$^\circ$\\ 
\end{tabular}
\end{ruledtabular}
\end{table}
%
Due to this large CP violation, the coupling of the lightest state $\phi_1$ 
to vector bosons can go down drastically for some values of $m_{H^+}$ 
and $\tan\beta$, see Fig.~\ref{cpxfig}(a). This results in highly suppressed 
production rates for the lightest Higgs boson; the lower bound on the 
mass of such a Higgs boson from direct searches at LEP can be as low as 
10--50 GeV~\cite{Abbiendi:2004ww}.
We take into account all three neutral Higgs bosons in the calculation of the 
fermion polarization by adding their $s$-channel diagrams. This may, however,
not be valid in some regions of the CPV-MSSM parameter space where the 
Higgs masses become nearly degenerate, i.e. the mass difference of two Higgs
bosons is smaller than sum of their widths and hence mixing between these 
states is resonantly enhanced. In this case one should do a coupled 
channel analyze~\cite{Choi:2004kq,Pilaftsis:1997dr} of the degenerate states. 
Fortunately, in the region of parameter space we consider, the mass differences
are always much larger than the sum of the decay widths, see 
Fig.~\ref{cpxfig}(c).  Thus our analysis is complementary to that of 
Refs.~\cite{Choi:2004kq,Pilaftsis:1997dr}.

\section{Numerical Results}
\label{five}

\begin{figure*}
\begin{tabular}{cc}
~{\tt CPsuperH}, $\Phi_{t,b,\tau}=0^\circ$ & 
~{\tt FeynHiggs}, $\Phi_{t,b,\tau}=0^\circ$\\
\epsfig{file=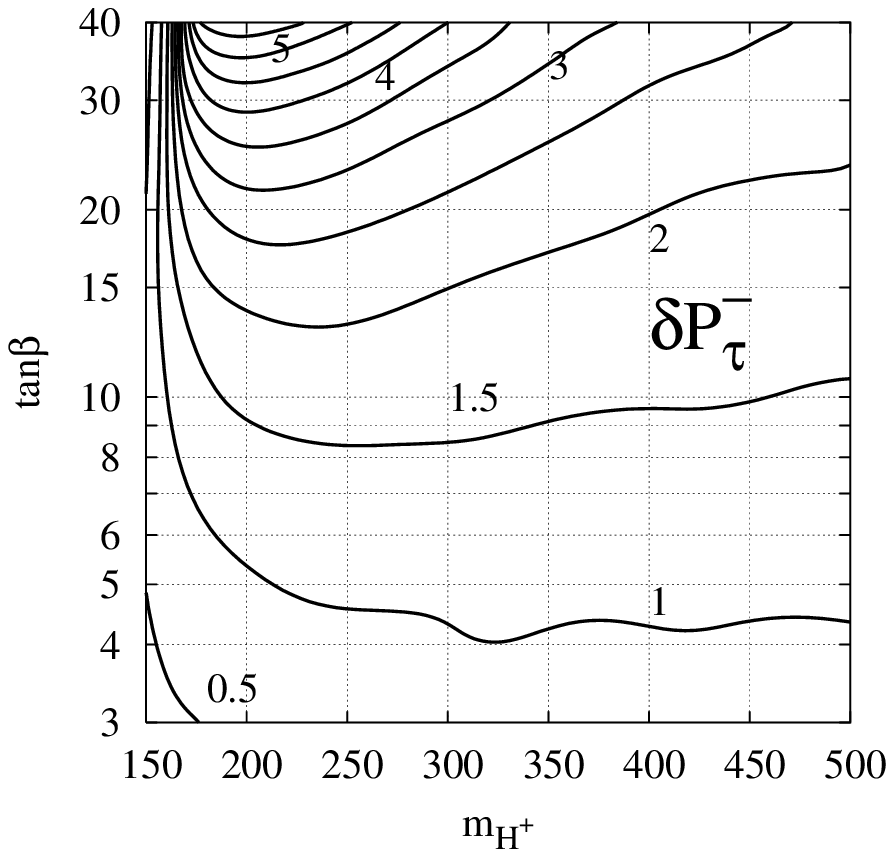,width=8.0cm}&
\epsfig{file=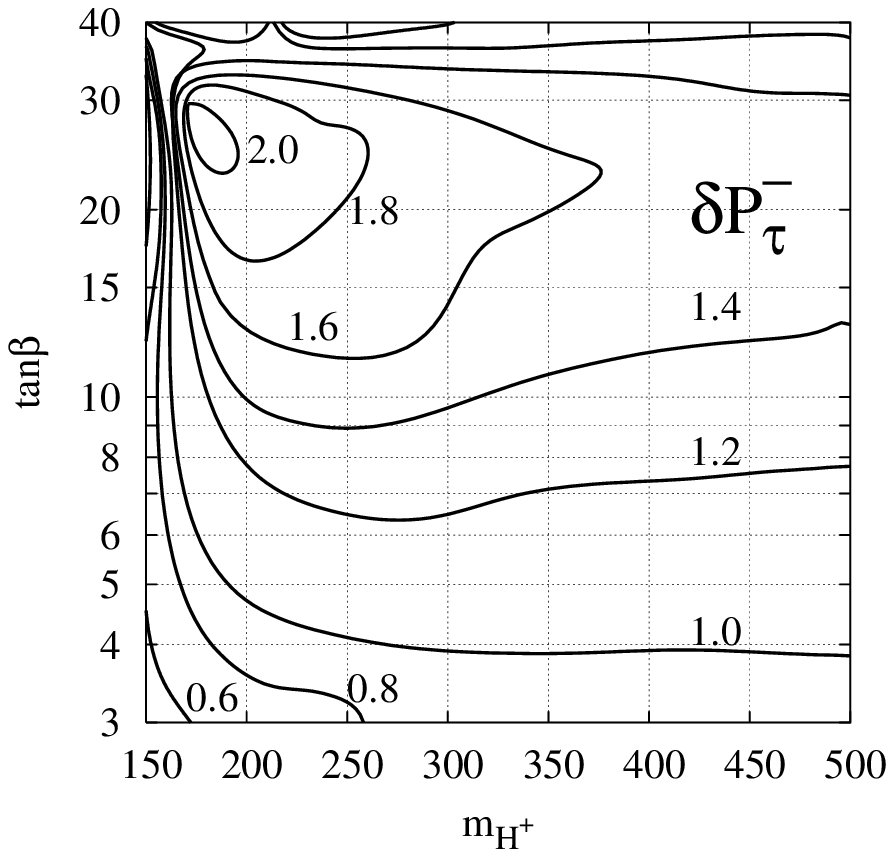,width=8.0cm}\\[2mm]
~{\tt CPsuperH}, $\Phi_{t,b,\tau}=90^\circ$ & 
~{\tt FeynHiggs}, $\Phi_{t,b,\tau}=90^\circ$\\
\epsfig{file=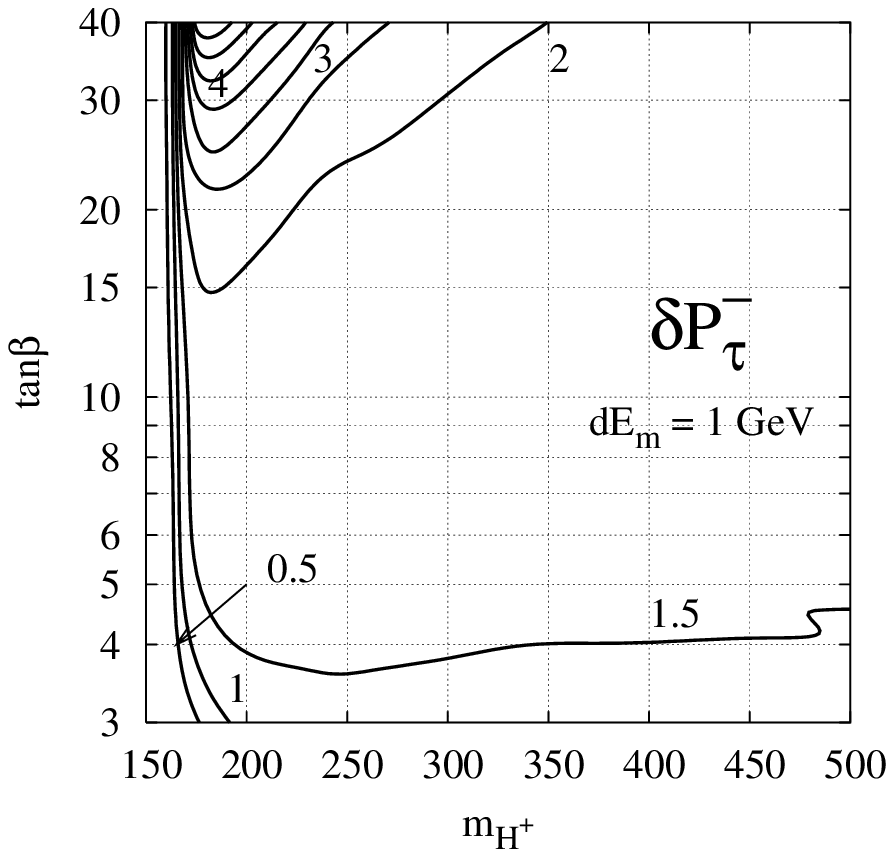,width=8.0cm} &
\epsfig{file=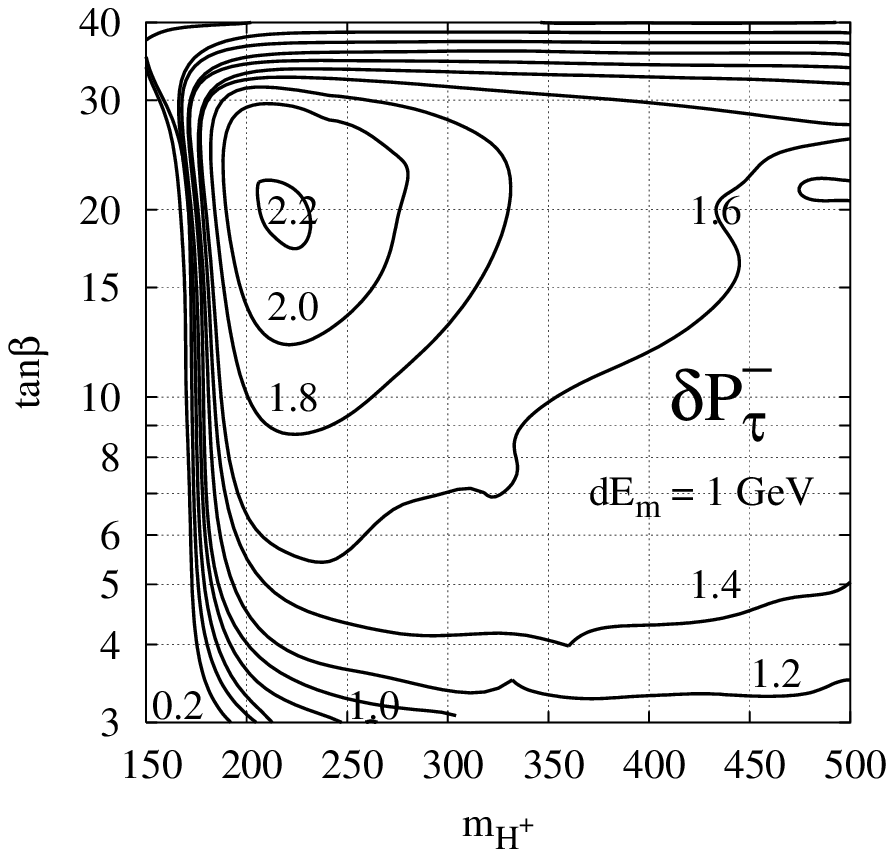,width=8.0cm}\\
\end{tabular}
\caption{\label{fig:tau-peak}Contours of constant $\delta P_\tau^-$ 
in units of $10^{-2}$ in the $(\tan\beta$--$m_{H^+})$ plane for the 
CPX scenario with $\Phi_{t,b,\tau}=0^\circ$ (top panels) and 
$\Phi_{t,b,\tau}=90^\circ$ (bottom panels) for the ``peak $E_b$" choice 
with $dE_m=1$~GeV. The left panels show the results obtained with 
{\tt CPsuperH}, the rights panels those obtained with {\tt FeynHiggs}.}
\end{figure*}
\begin{figure*}
\begin{tabular}{cc}
~{\tt CPsuperH}, $\Phi_{t,b,\tau}=0^\circ$ & 
~{\tt FeynHiggs}, $\Phi_{t,b,\tau}=0^\circ$\\
\epsfig{file=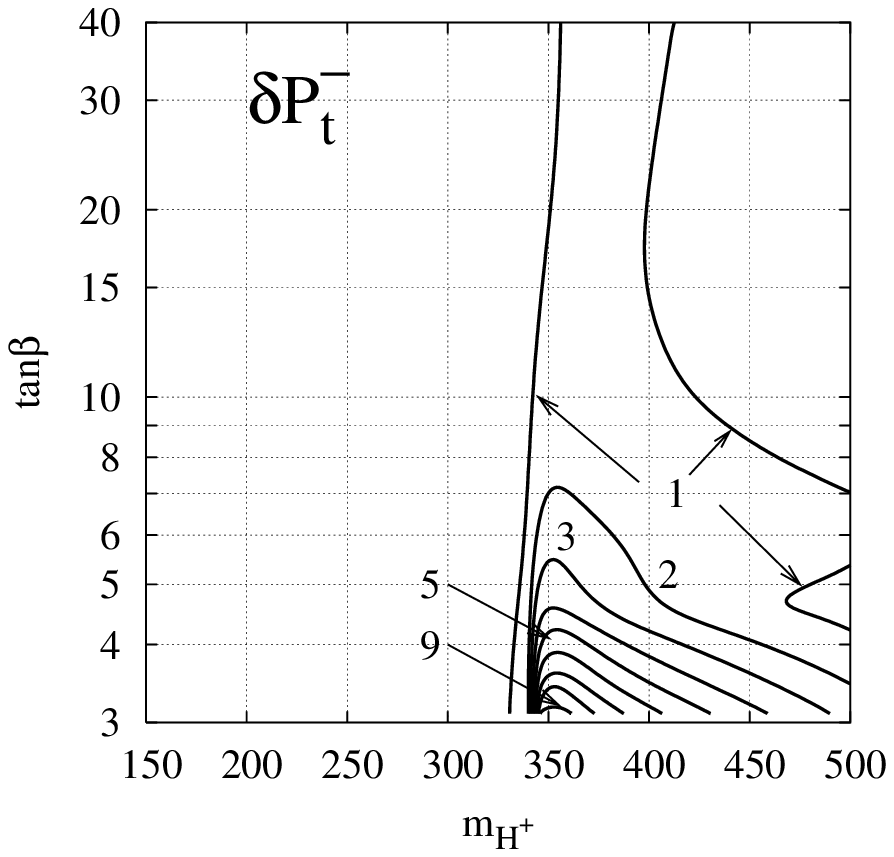,width=8.0cm} &
\epsfig{file=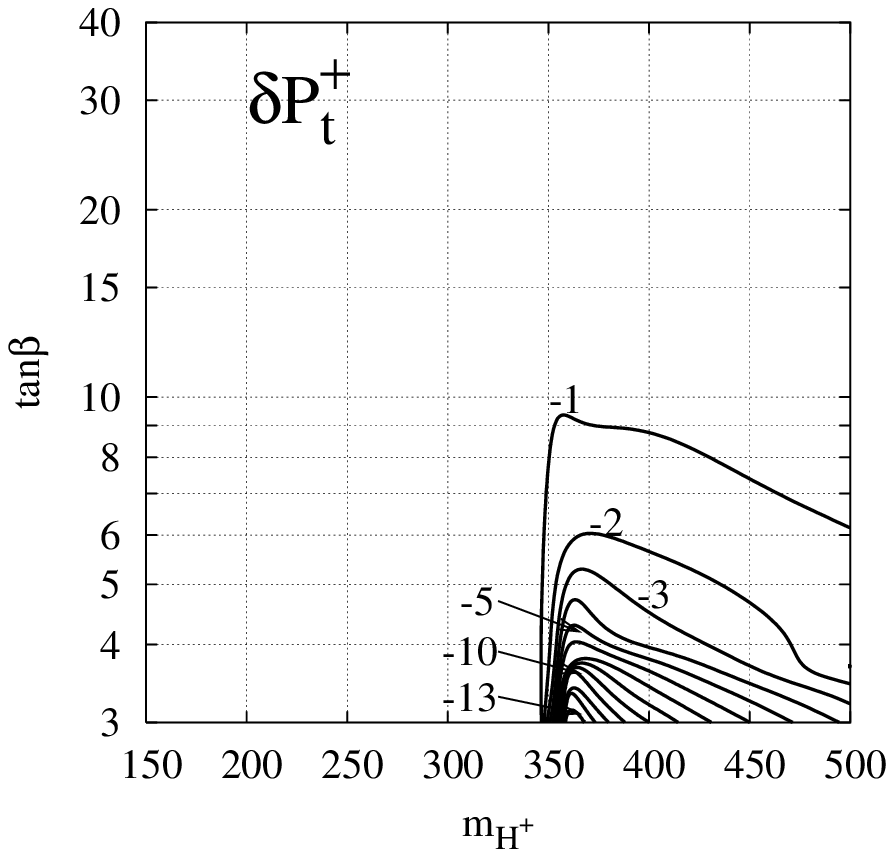,width=8.0cm}\\[2mm]
~{\tt CPsuperH}, $\Phi_{t,b,\tau}=90^\circ$ & 
~{\tt FeynHiggs}, $\Phi_{t,b,\tau}=90^\circ$\\
\epsfig{file=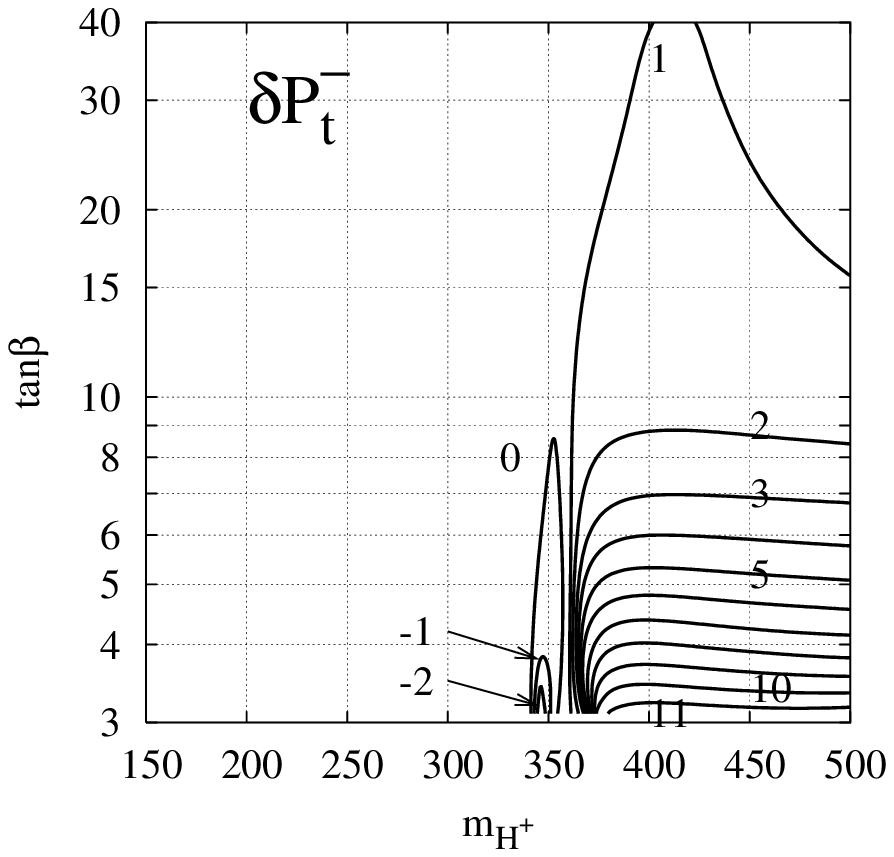,width=8.0cm} &
\epsfig{file=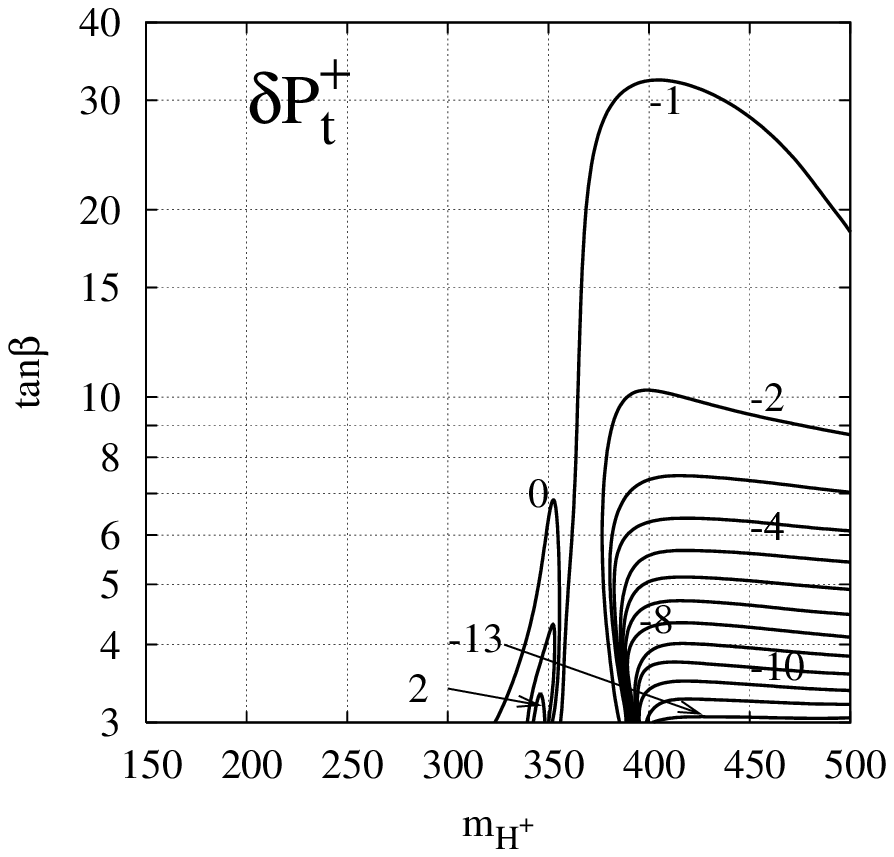,width=8.0cm}\\\
\end{tabular}
\caption{\label{fig:top-peak}Contours of constant $\delta P_t^\pm$ 
in units of $10^{-2}$ in the $(\tan\beta$--$m_{H^+})$ plane for the 
CPX scenario with $\Phi_{t,b,\tau}=0^\circ$ (top panels) and 
$\Phi_{t,b,\tau}=90^\circ$ (bottom panels) for the ``peak $E_b$" choice. 
The left panels show the results for $\delta P_t^-$ 
obtained with {\tt CPsuperH} and  the rights panels those for
$\delta P_t^+$ obtained with {\tt FeynHiggs}, see text.}
\end{figure*}
At a photon collider, the center of mass
energy of the colliding photons is not fixed 
but has a wide spectrum. The shape of this spectrum depends upon the 
polarizations of the laser and the $e^+/e^-$ beams of the parent collider.
For the numerical calculation of cross sections, we use the ideal Compton 
back-scattered spectrum~\cite{ginzburg} with polarizations of the electron 
beam and the laser chosen such that one obtains the hard spectrum
in Fig~\ref{spec}. 
The parent electron beam energy, $E_b$, can be chosen to maximize the 
deviation from the QED prediction, cf. Fig.~\ref{p-eb}. Our 
observables are maximized when the Higgs mass matches the value of the 
$\gamma\gamma$ invariant mass at the peak of the polarized photon spectrum. 
This happens  for  $E_b=(m_{\phi_i}/2)/z$, where $z$ (the scaled $\gamma\gamma$
invariant mass, $z=\sqrt{\omega_1\omega_2}/E_b$) takes a value between 0.75 to 
0.83 for $x_c = 4.8$.  This corresponds to the case where the scaled
Higgs mass, $m_{\phi_i}/2E_b$, lies in the  grey band of  Fig.~\ref{spec}. 

We can therefore pursue two different strategies for choosing $E_b$ :
\begin{enumerate}
  \item Parameterizing the relationship between $m_{\phi_i}$ and $E_b$ 
        in terms of $z_0$ as $E_b = (m_{\phi_i}/2)/z_0$, we choose an
	optimal value of $z_0$, say $z_0=0.80$, for each point in the 
	scan such that $\delta P$ is maximized. This gives a very good 
	estimate of the ultimate potential of the particular observable 
	used in adaptation. We call this the ``peak $E_b$" choice.
  \item Fixing $E_b$ such that the relevant  Higgs mass ($m_{\phi_1}$ 
        for $\tau^+ \tau^-$ and $m_{\phi_2, \phi_3}$ for $t \bar t$ production)
	  matches approximately
        with the  $\gamma \gamma$ invariant mass corresponding to $z$ values
        within the peak of the photon spectrum (the grey band in 
	  Fig.~\ref{spec}). Though this choice does not exploit the 
        observable optimally, it is closer to what will be done in a 
        realistic experiment. We call this the ``fixed $E_b$" choice.
\end{enumerate}
In the case of $\tau$ polarization,
due to the small values of $m_{\tau}$ and $\Gamma_{\phi_1}$, the absolute
values of the polarization observables are $\lsim 10^{-5}$. These can be 
enhanced by putting a cut on the invariant mass of the $\tau^+ \tau^-$ pair
to select the ones coming from $\phi_1$ decay~\cite{asakawa}~:
\begin{equation}
 |m_{\tau\tau}-m_{\phi_1}| \leq \max(dE_m, \ 5 \ \Gamma_{\phi_1}),
\end{equation}
where $dE_m$ is the minimum resolution of 
$m_{\tau\tau}$ reconstruction. We use  $dE_m=$~1~GeV for purposes of
illustration in this paper. For the case of top production
such a cut is not necessary.

In the following, we perform a scan over the MSSM parameters as given in
Table~\ref{mspara} and calculate the $\tau$ and $t$ polarization observables
for both the peak and the fixed $E_b$ choices using both 
{\tt CPsuperH} and {\tt FeynHiggs}
for calculating the Higgs masses, couplings, and widths.
The statistical fluctuation in the value of the fermion polarization 
is given by
\begin{equation}
\Delta P_f = \frac{\sqrt{1-P_f^2}}{\sqrt{N_++N_-}}\,.
\end{equation}
The typical value of $\Delta P_f$ for an integrated luminosity of 
100~fb$^{-1}$ is about 0.003 for a total rate of 1~pb. The typical $\tau$-pair
production rate with the above cut is about 1--10~pb over the
$(\tan\beta$--$m_{H^+})$ plane. $\tau$ decays into
hadronic channels reduce the useful cross section and hence
cause this error to increase. As a conservative measure we take 
$\delta P_\tau \ge 0.01$ in order to be measurable. 
The $t\bar t$ production rate, on the other hand, is less than 1~pb for the
energies considered, and it can be as low as 8~fb with the ``peak $E_b$" 
choice in some regions of the $(\tan\beta$--$m_{H^+})$ plane. 
Thus the statistical error in the polarization measurement 
goes up and the sensitivity goes down in this case, even if 
the polarization asymmetry is large.

\subsection{Peak $E_b$ scan}

\begin{figure*}
\begin{tabular}{cc}
~{\tt CPsuperH}, $\Phi_{t,b,\tau}=90^\circ$ & 
~{\tt FeynHiggs} $\Phi_{t,b,\tau}=90^\circ$\\
\epsfig{file=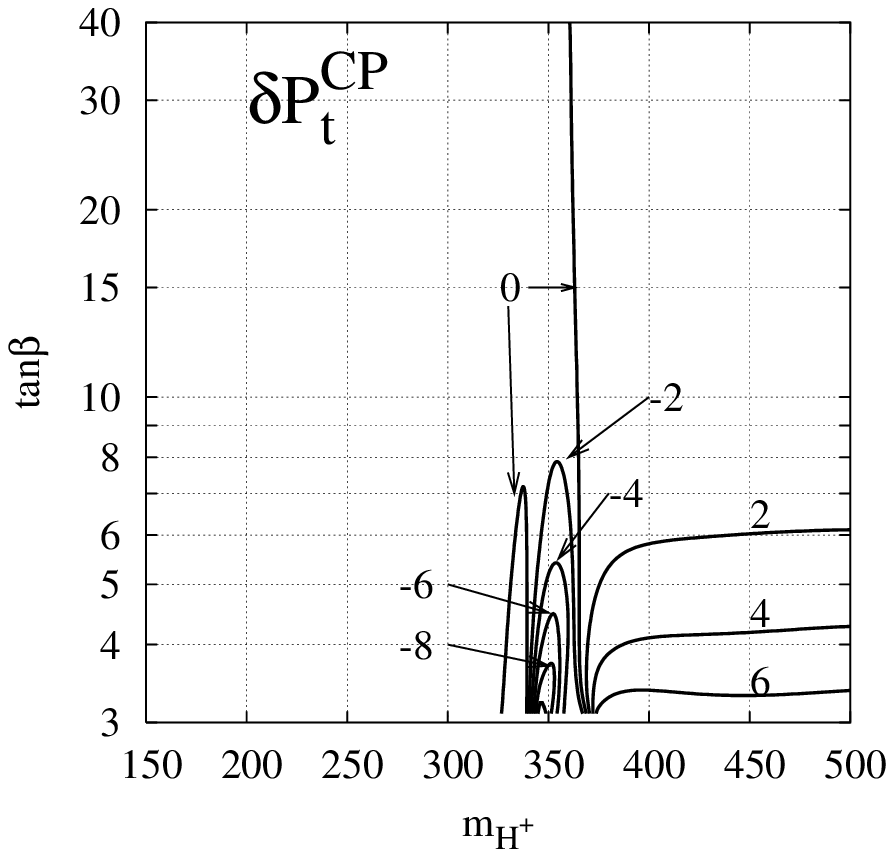,width=8.0cm} &
\epsfig{file=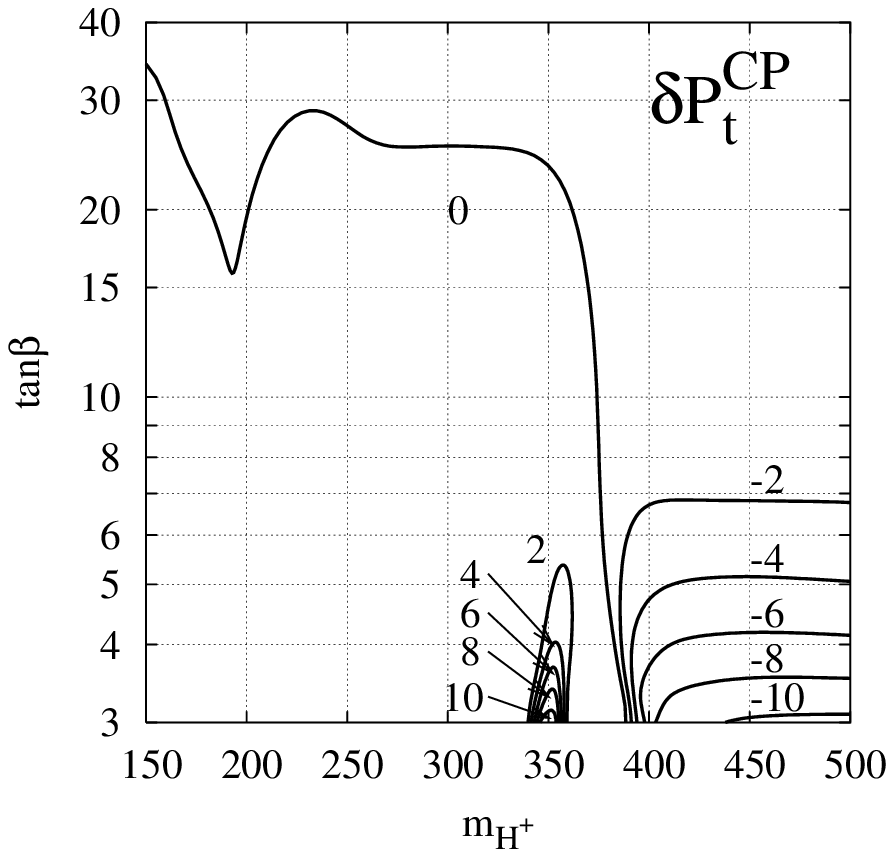,width=8.0cm}\\
\end{tabular}
\caption{\label{fig:top-peak-CP}Contours of constant $\delta P_t^{\rm CP}$ 
in units of $10^{-2}$ in the $(\tan\beta$--$m_{H^+})$ plane for the 
CPX scenario with $\Phi_{t,b,\tau}=90^\circ$ and ``peak $E_b$" choice.
The left panel shows the results obtained with 
{\tt CPsuperH}, the right panel those obtained with {\tt FeynHiggs}.}
\end{figure*}

We first discuss the case of $\tau$-pair production through $\phi_1$ exchange
for the ``peak $E_b$" choice. Figure~\ref{fig:tau-peak} shows contours of
constant $\delta
P^-_\tau$ as obtained with {\tt CPsuperH} and {\tt FeynHiggs} in the
($\tan\beta-m_{H^+}$) plane for $\Phi_{t,b,\tau}=0^\circ$ and
$\Phi_{t,b,\tau}=90^\circ$. For each point in the scan, the beam energy is set
to $E_b=m_{\phi_1}/(2z_0)$ to maximize $\delta P^\pm_\tau$. In the
CP-conserving case, $\Phi_{t,b,\tau}=0^\circ$, $\delta P_\tau^-$ should be
measurable for $\tan\beta \ge 4$ and $m_{H^+}\ge 200$ GeV; smaller values of
$m_{H^+}$ require somewhat higher $\tan\beta$ to achieve $\delta P_\tau^- \ge
0.01$. In the case of maximal CPV phases, $\Phi_{t,b,\tau}=90^\circ$, $\delta
P_\tau^- > 0.01$ holds over practically the entire parameter range, including
the region where a very light Higgs boson may have been missed at 
LEP2~\cite{Abbiendi:2004ww}.
Such a light CPV Higgs boson will also be difficult to discover at the 
LHC~\cite{schumacher}.
The process $\gamma\gamma \to \tau^+\tau^-$ may hence offer a unique
possibility for this case. While $\delta P_\tau^-$ covers a large part of the
parameter space, the CP-odd observable $\delta P_\tau^{\rm CP}$ is very small,
$\delta P_\tau^{\rm CP} \sim 10^{-5}$, and hence below the limit of  
measurability even if the 
CPV phases are maximal. This means in turn that $\delta P^+_\tau\approx-\delta
P^-_\tau$. It is also worth noting that the size of the observable is rather
sensitive to the value of $dE_m$~: increasing for instance $dE_m$ from 1 GeV 
to 2 GeV, $\delta P^-_\tau$ goes down by about a factor of 2 over most of the
parameter space in Fig.~\ref{fig:tau-peak}.

Let us now turn to the top polarization in $\gamma\gamma\to t\bar t$. Due to
the large top quark mass, here only the heavier Higgs bosons $\phi_{2,3}$
contribute. For the comparatively large $m_{H^+}$ values needed to obtain
$m_{\phi_{2,3}} \ge 2m_t$, the mass difference between $\phi_2$ and $\phi_3$ is
usually so small that both their scaled masses can be within the peak of the
hard photon spectrum of Fig.~\ref{spec}, although 
$m_{\phi_3}-m_{\phi_2}~\gg~\Gamma_{2,3}$ is still maintained. 
We therefore choose
\begin{equation}
 E_b^{peak} = \max\left[\, E_b^0, (m_{\phi_2}+m_{\phi_3})/(4z_0)\, \right]\,,
\end{equation}
where $E_b^0=220$ GeV and $z_0=0.8$. Figure~\ref{fig:top-peak} shows contours 
of constant $\delta P_t^\pm$ in the ($\tan\beta-m_{H^+}$) plane analogous to 
Fig.~\ref{fig:tau-peak}. Owing to different sign conventions the two programs, 
$\delta P_t^\pm$ of {\tt CPsuperH} corresponds to $-\delta P_t^\mp$ in {\tt
FeynHiggs}. In Fig.~\ref{fig:top-peak} we therefore plot $\delta P_t^-$ 
for {\tt CPsuperH} and $\delta P^+_t$ for {\tt FeynHiggs}. 
The deviation from the pure QED prediction
(i.e.\ the Higgs contribution) is sizable for $\tan\beta \lsim 10$. Note that 
the CP-even polarization observables for $t\bar t$ can reach much larger values
than those for $\tau^+\tau^-$. Moreover, in the case of $\gamma\gamma \to t
\bar t$ also the CP-odd observable, $\delta P_t^{\rm CP}$, may be large enough to
be measurable. This is shown in Fig.~\ref{fig:top-peak-CP} for
$\Phi_{t,b,\tau}=90^\circ$.
Because of the low cross sections, ranging from 8 fb to 150 fb in 
Fig.~\ref{fig:top-peak-CP}, as compared to $1$--$10$ pb for the $\tau$ case, 
the statistical fluctuations are large, about $\Delta P_t~\sim 0.10 - 0.03$. 
Therefore the region of sensitivity to the Higgs-boson contributions is 
restricted to low $\tan\beta$ values.

\subsection{Fixed $E_b$ scan}
\begin{figure*}
\begin{tabular}{cc}
{\tt CPsuperH} $\Phi_{t,b,\tau}=0^\circ$ & 
{\tt FeynHiggs} $\Phi_{t,b,\tau}=0^\circ$\\
\epsfig{file=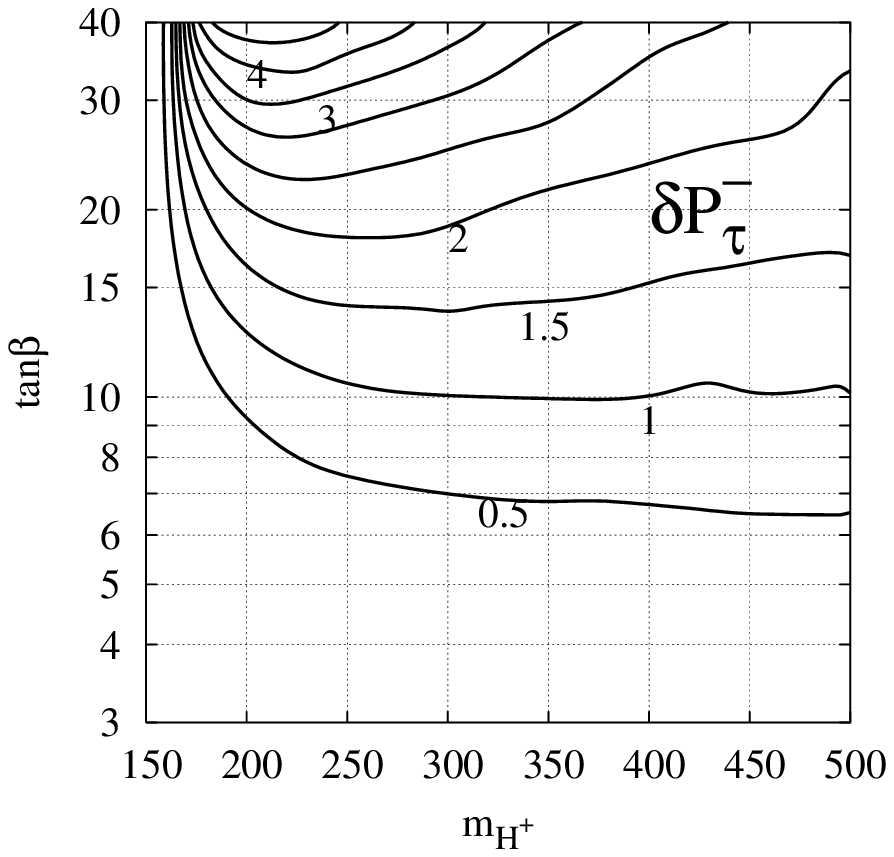,width=8.0cm}&
\epsfig{file=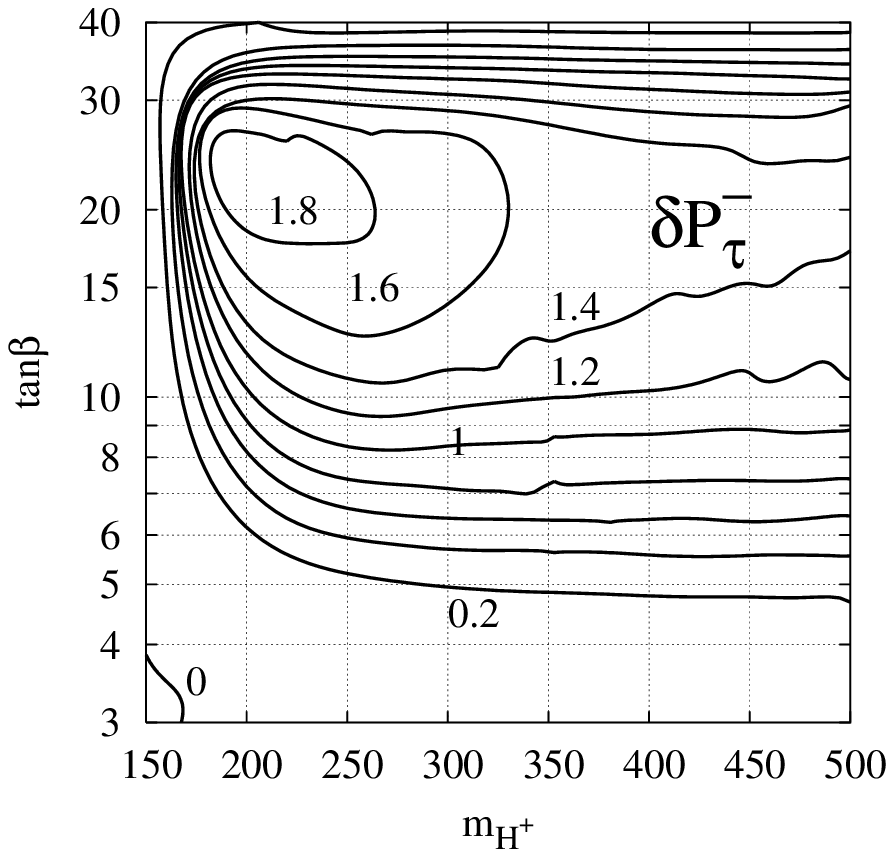,width=8.0cm}\\
{\tt CPsuperH} $\Phi_{t,b,\tau}=90^\circ$ & 
{\tt FeynHiggs} $\Phi_{t,b,\tau}=90^\circ$\\
\epsfig{file=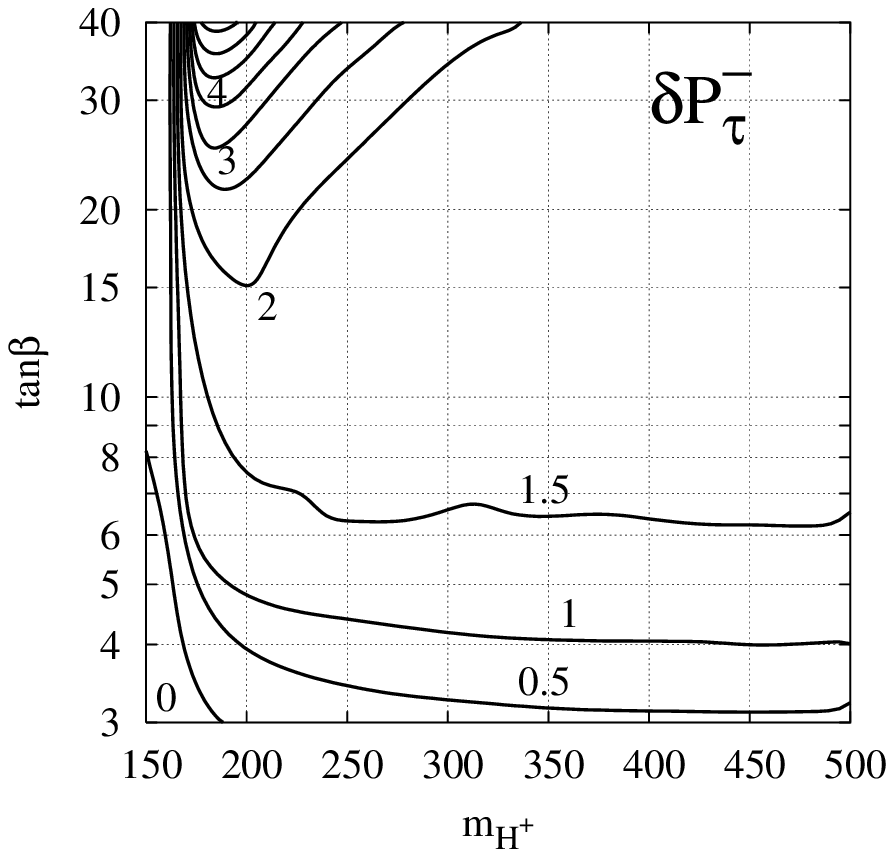,width=8.0cm}&
\epsfig{file=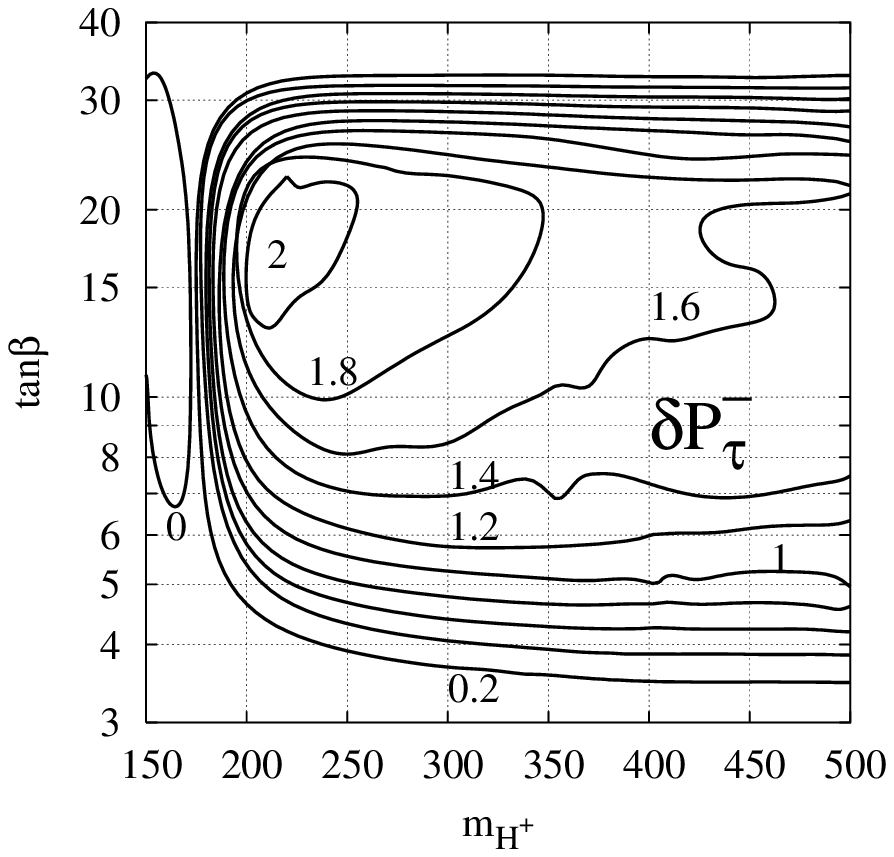,width=8.0cm}
\end{tabular}
\caption{\label{fig:tau-fix-ch}Contours of constant $\delta P_\tau^-$ in 
units of $10^{-2}$ for
fixed $E_b$ and $\Phi_{t,b,\tau}=0^\circ$ and $90^\circ$ using {\tt CPsuperH}
(left panels) and {\tt FeynHiggs} (right panels) to compute the Higgs masses,
couplings and widths. $E_b=77$~GeV, except for the
lower-right plot where $E_b=82$~GeV, see text.}
\end{figure*}
\begin{figure*}
\begin{tabular}{cc}
~{\tt CPsuperH} &
~{\tt FeynHiggs} \\
\epsfig{file=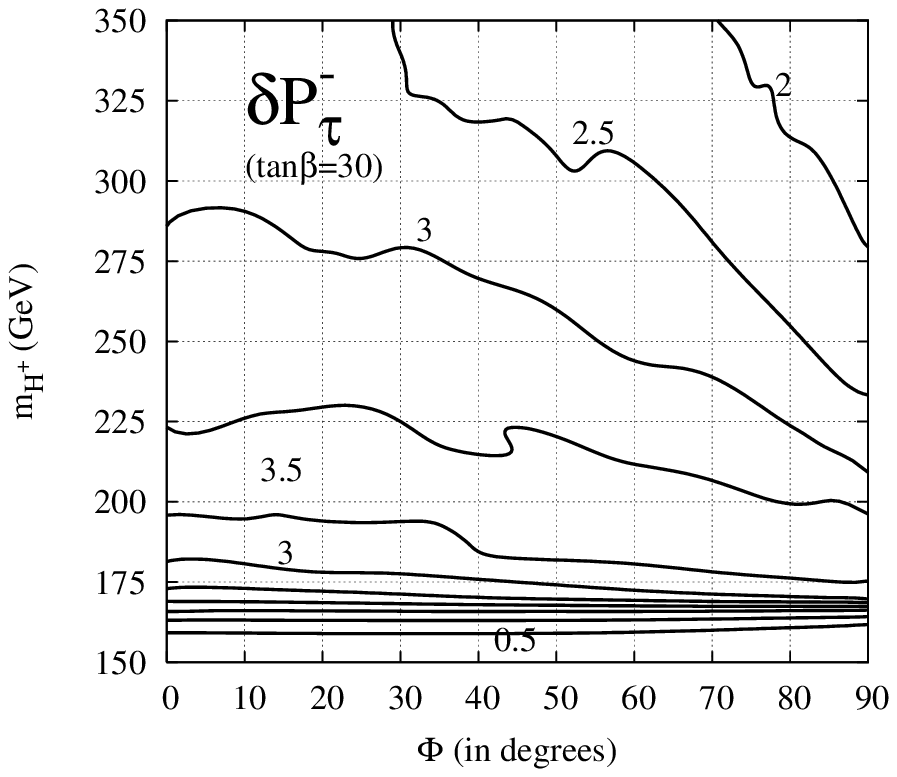,width=8.0cm}&
\epsfig{file=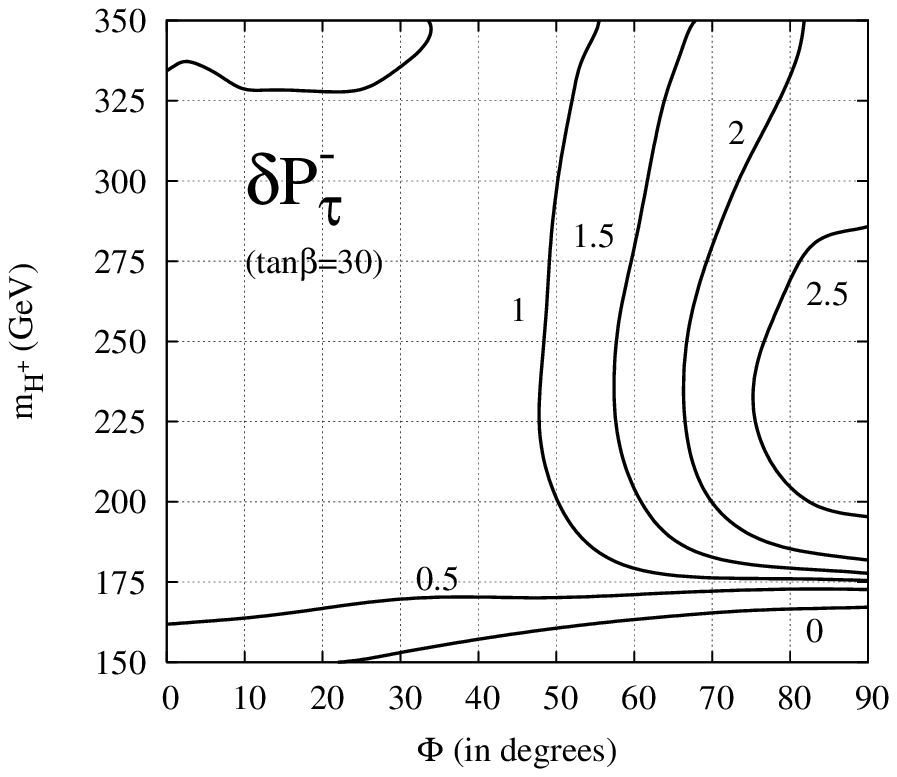,width=8.0cm}
\end{tabular}
\caption{\label{fig:tau-phi}Contours of constant $\delta P_\tau^-$ in units 
of $10^{-2}$  in the
($\Phi_{t,b,\tau}-m_{H^+}$) plane for $\tan\beta=30$ and $E_b=77$~GeV
 with {\tt CPsuperH} (left panel) and {\tt FeynHiggs} (right panel).}
\end{figure*}
\begin{figure*}
\begin{tabular}{cc}
{\tt CPsuperH} $\Phi_{t,b,\tau}=0^\circ$ & 
{\tt FeynHiggs} $\Phi_{t,b,\tau}=0^\circ$\\
\epsfig{file=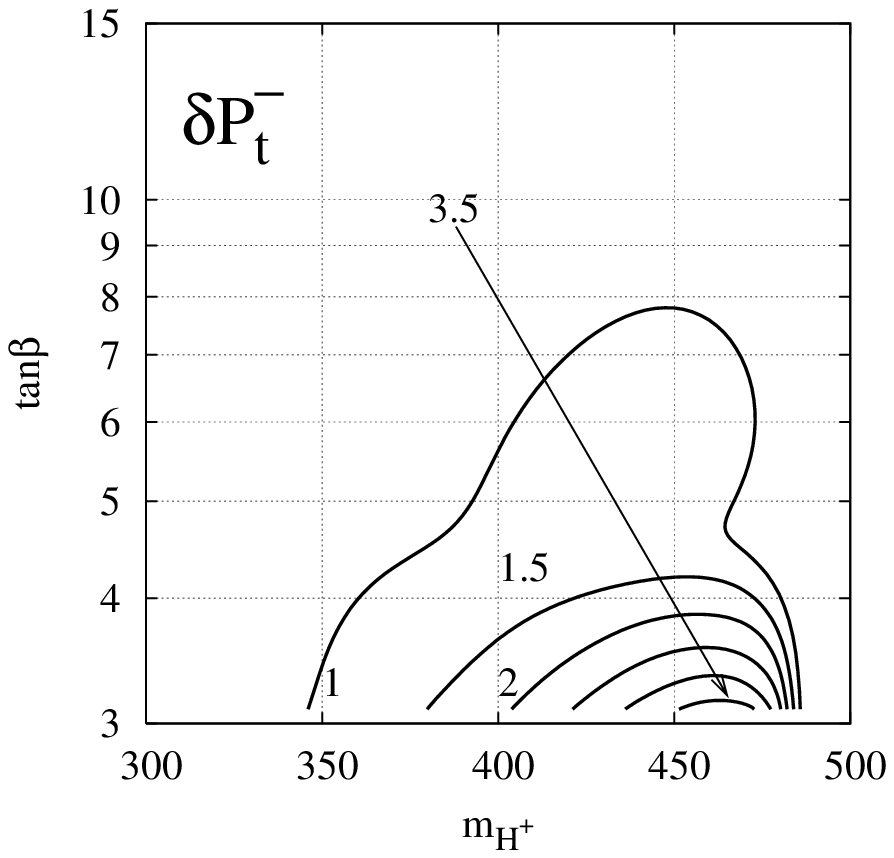,width=8.0cm}&
\epsfig{file=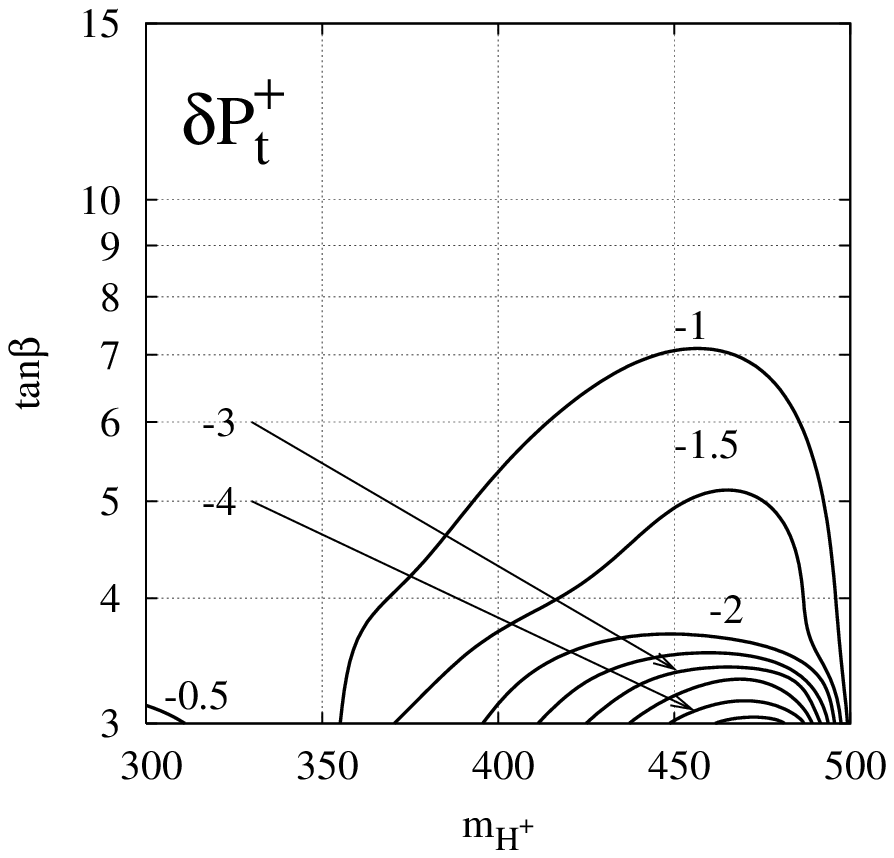,width=8.0cm}\\
{\tt CPsuperH} $\Phi_{t,b,\tau}=90^\circ$ & 
{\tt FeynHiggs} $\Phi_{t,b,\tau}=90^\circ$\\
\epsfig{file=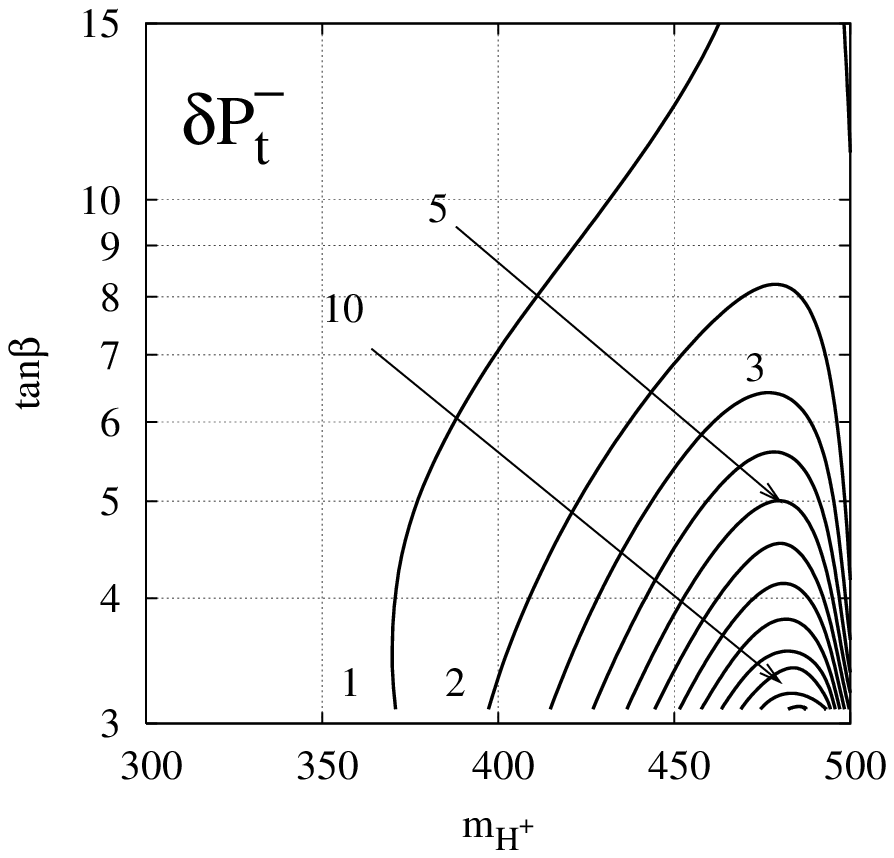,width=8.0cm}&
\epsfig{file=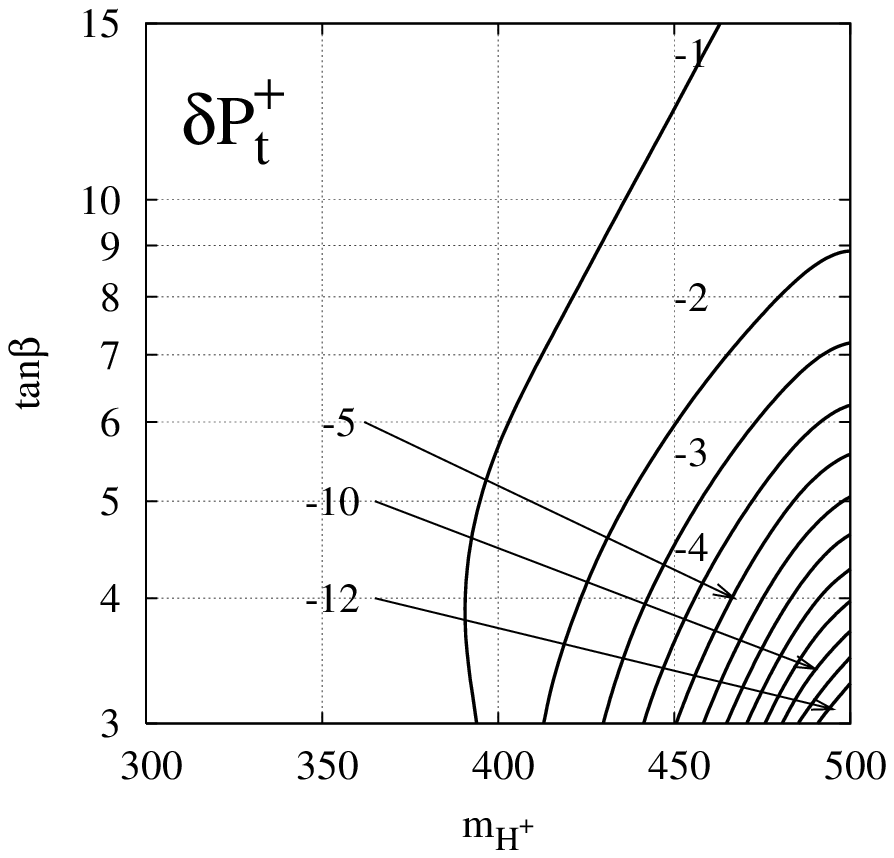,width=8.0cm}
\end{tabular}
\caption{\label{fig:top-fix}Contours of constant $\delta P_t^\pm$ in units 
of $10^{-2}$ for fixed $E_b=300$ GeV and $\Phi_{t,b,\tau}=0^\circ$ (top) 
and $90^\circ$ (bottom) using {\tt CPsuperH} (left) and {\tt FeynHiggs} (right)
to compute the Higgs masses, couplings and widths.}
\end{figure*}
The ``peak $E_b$" choice discussed above gives an estimate of the ultimate
potential of our polarization observables. In reality, however, one will have a
collider running at some fixed beam energy. Obviously it will be of advantage
to set $E_b$ such that one is sensitive to the Higgs contributions over a large
part of the parameter space.

In the CPX scenario, {\tt CPsuperH} predicts $m_{\phi_1}\lsim 123$ GeV. Thus 
$E_b=77$~GeV leads to a good sensitivity over most of the parameter space.
With {\tt FeynHiggs}, however, the maximum value of $m_{\phi_1}$ considerably
changes with $\Phi_{t,b,\tau}$ in the scan; we obtain maximum
values of $m_{\phi_1}=$ 123 GeV for $\Phi_{t,b,\tau}=0^\circ$ and $m_{\phi_1}=$
131 GeV for $\Phi_{t,b,\tau}=90^\circ$, respectively. Hence we choose $E_b=77$~GeV
for $\Phi_{t,b,\tau}=0^\circ$ and $E_b=82$~GeV for $\Phi_{t,b,\tau}=90^\circ$ 
in the computation with {\tt FeynHiggs}. 
The polarization observable $\delta P_\tau^-$ for this fixed
$E_b$ choice is shown in the ($\tan\beta-m_{H^+}$) plane in 
Fig.~\ref{fig:tau-fix-ch}. We observe
that for $\Phi_{t,b,\tau}=90^\circ$, $\delta P_\tau^- \ge 0.01$ unless
$\tan\beta$ is very small. For $\Phi_{t,b,\tau}=0^\circ$, on the other hand,
observability of $\delta P_\tau^-$ is limited to $\tan\beta \gsim 8$--10. It is
apparent that $\delta P_\tau$ will be mainly useful if $\tan\beta$ is large. To
explicitly see the phase dependence we show in 
Fig.~\ref{fig:tau-phi} contours of constant
$\delta P^-_\tau$ in the ($m_{H^+}-\Phi$) plane for $E_b=77$ GeV. There is a
rather large difference in the results of the two codes, which is also apparent
in the other figures, due to differences in the implementation of radiative
corrections in the two programs~\cite{sven}. It is clear that for analyses as 
suggested in this paper, more precise computations will be 
necessary.

\begin{figure*}
\begin{tabular}{cc}
~{\tt CPsuperH} &
~{\tt FeynHiggs}\\
\epsfig{file=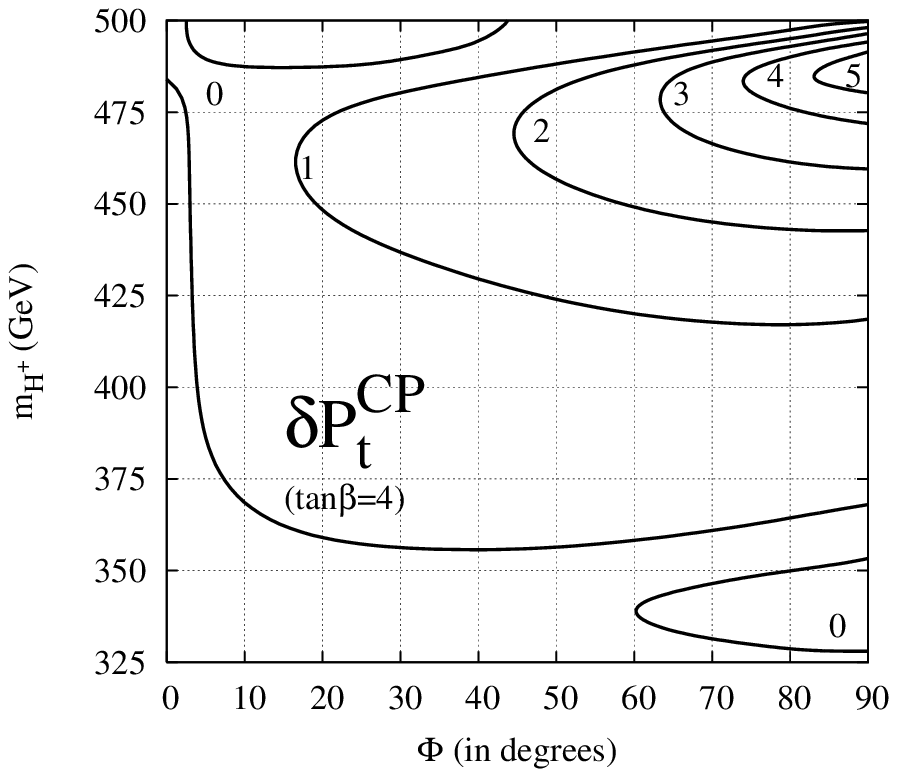,width=8.0cm}&
\epsfig{file=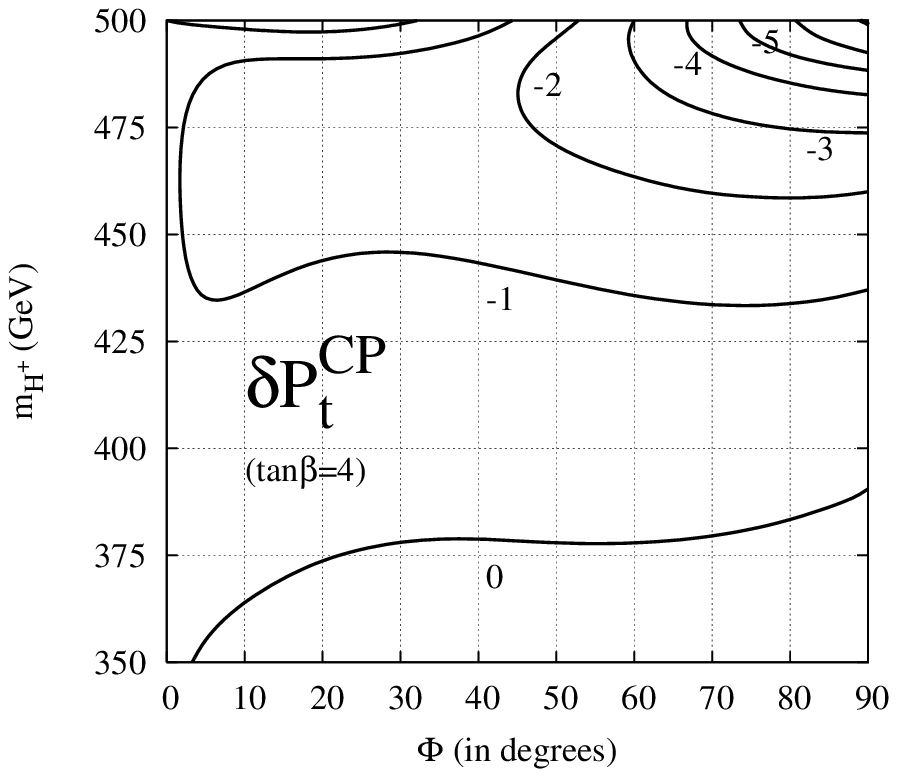,width=8.0cm}
\end{tabular}
\caption{\label{fig:top-phim}Contours of constant $\delta P_t^{\rm CP}$ in
units of $10^{-2}$ in
the ($\Phi_{t,b\tau}-m_{H^+}$) plane for $\tan\beta=4$ and $E_b=300$~GeV  
with {\tt CPsuperH} (left panel) and {\tt FeynHiggs} (right panel).}
\end{figure*}

\begin{figure*}
\begin{tabular}{cc}
~{\tt CPsuperH} &
~{\tt FeynHiggs}\\
\epsfig{file=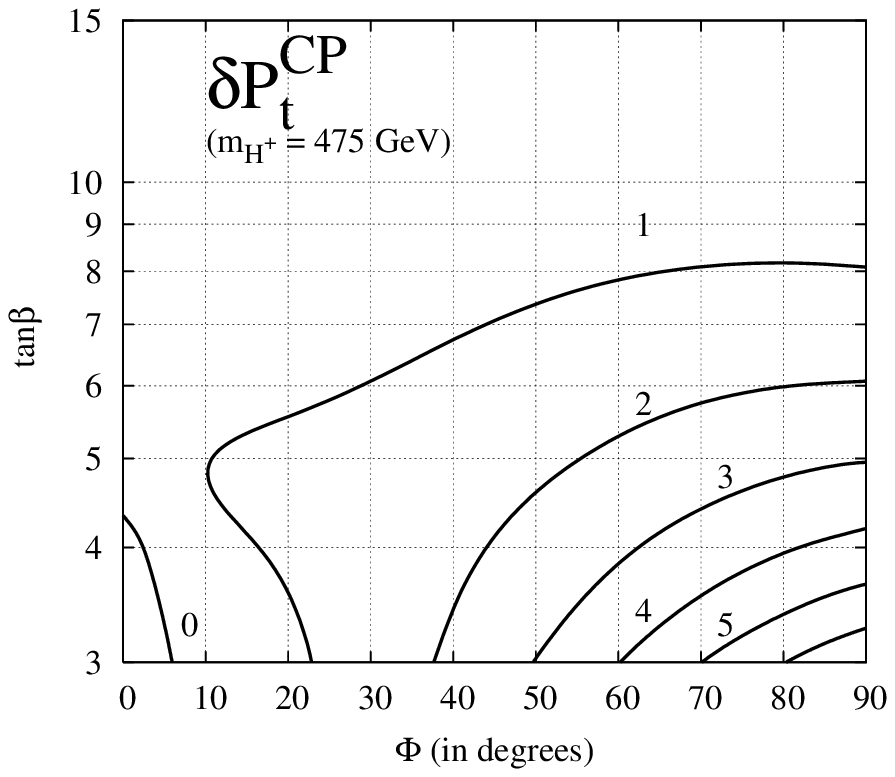,width=8.0cm}&
\epsfig{file=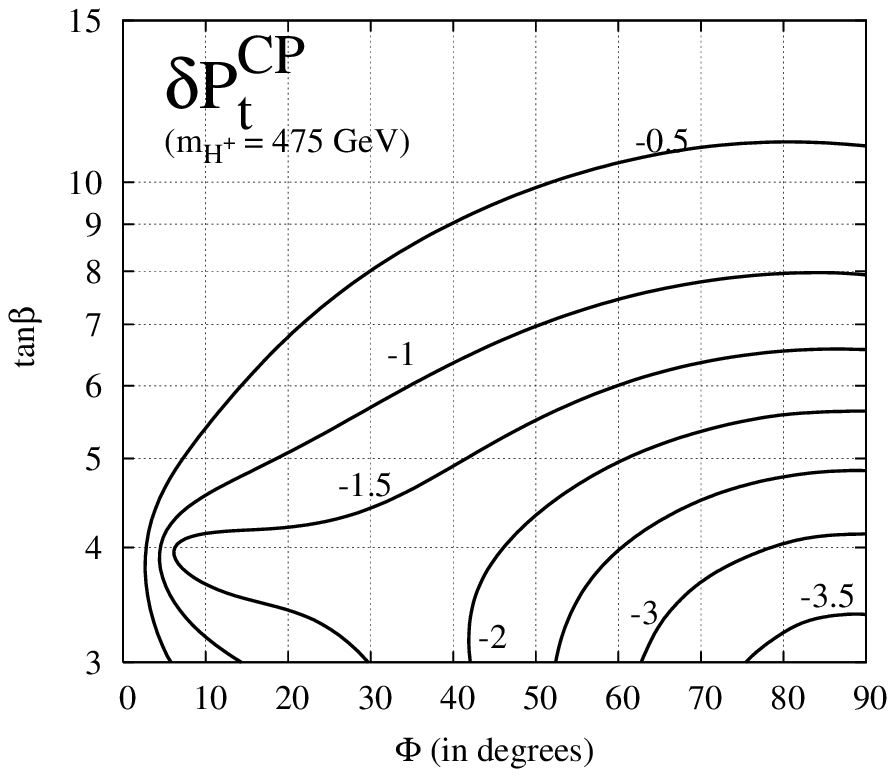,width=8.0cm}
\end{tabular}
\caption{\label{fig:top-phit}Contours of constant $\delta P_t^{\rm CP}$ in 
units of $10^{-2}$ in
the ($\Phi_{t,b\tau}-\tan\beta$) plane for $m_{H^+}=475$ GeV and $E_b=300$~GeV
with {\tt CPsuperH} (left panel) and {\tt FeynHiggs} (right panel).}
\end{figure*}

We next turn to $t\bar t$ production with fixed $E_b$. In this case, as for
the peak $E_b$ choice, it is the mean mass of $\phi_2$ and $\phi_3$ that should
be within the peak of the photon spectrum. However, since $m_{\phi_{2,3}}$
change linearly with $m_{H^+}$, one cannot have optimal sensitivity over the
whole parameter space with fixed $E_b$. We hence take $E_b=300$ GeV as a good
compromise. For this choice one has comparatively large rates while the scaled 
masses of $\phi_{2,3}$ still lie within the peak of the photon spectrum for a 
sizable portion of the ($\tan\beta-m_{H^+}$) plane. 
The results for $\delta P_t^\pm$ obtained with
{\tt CPsuperH} and {\tt FeynHiggs} for $\Phi_{t,b,\tau}=0^\circ$ and
$\Phi_{t,b,\tau}=90^\circ$ are shown in 
Fig.~\ref{fig:top-fix}. Again the role of
$\delta P_t^+$ and $\delta P^-_t$ is interchanged in the two codes. 
In Fig.~\ref{fig:top-phim},
we show contours of constant $\delta P_t^{\rm CP}$ in the ($\Phi-m_{H^+}$) plane
for $\tan\beta=4$ and in 
Fig.~\ref{fig:top-phit} in the ($\Phi-\tan\beta$) plane for
$m_{H^+}=475$ GeV. As one can see, there is sensitivity to CP violation if
$\tan\beta$ is small. Moreover, there is rather good agreement between 
{\tt CPsuperH} and {\tt FeynHiggs} in $\delta P_t^{\rm CP}$ (up to a sign).
Note also that the signal can be enhanced by tuning $E_b$. 

Comparing these results with the ``peak $E_b$" choice, we see that for 
$\gamma\gamma\to\tau\tau$ one can be sensitive to as large a region of 
the parameter space if $E_b$ is chosen carefully. 
For $\gamma\gamma\to t\bar t$, on the other hand, we loose 
sensitivity to part of the parameters space with fixed $E_b$.


\subsection{Lepton asymmetries}

\begin{figure*}
\begin{tabular}{cc}
$\Phi_{t,b,\tau}=30^\circ$&
$\Phi_{t,b,\tau}=90^\circ$\\
\epsfig{file=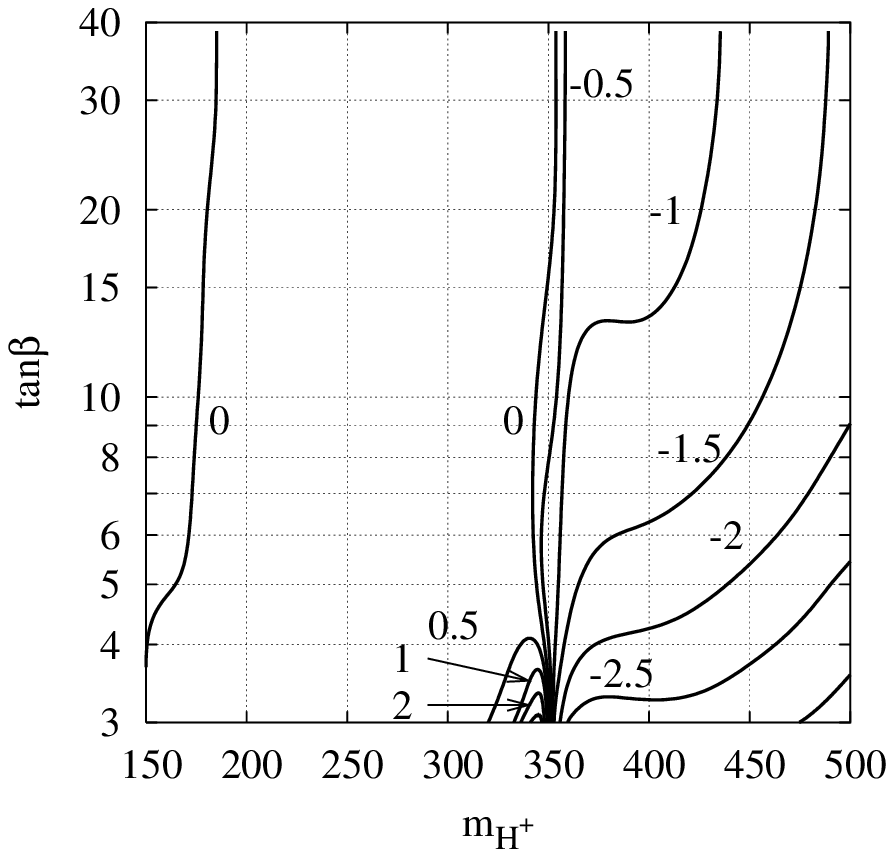,width=8.0cm}&
\epsfig{file=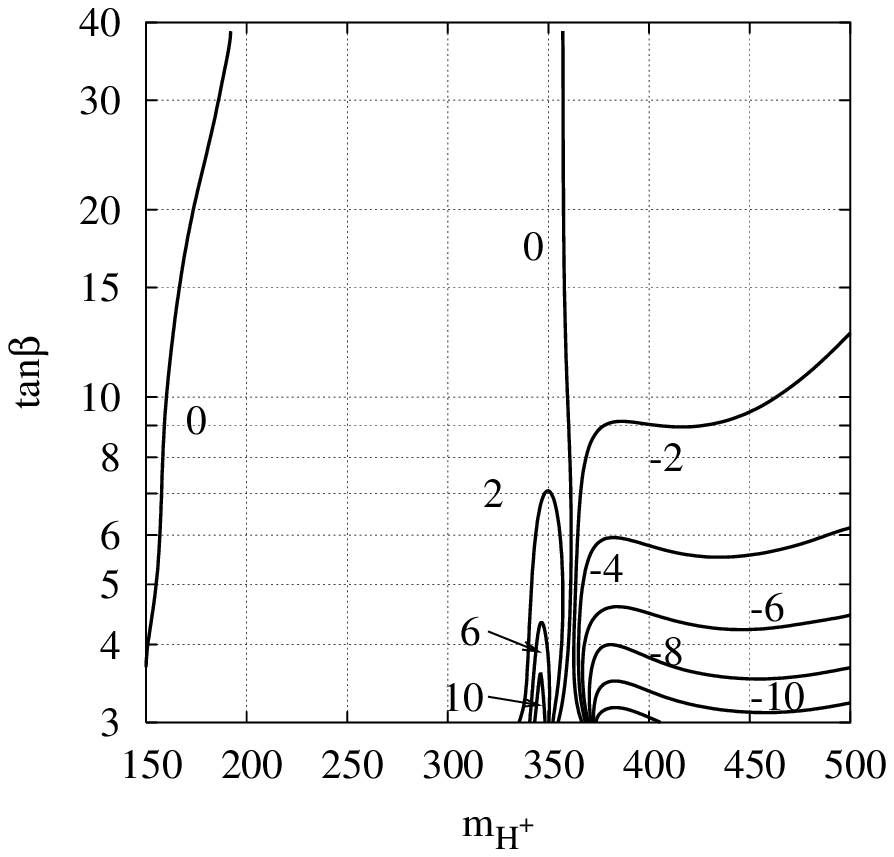,width=8.0cm} 
\end{tabular}
\caption{\label{fig:lept-peak-ch}Contours of constant ${\cal A}_3$ in units 
of $10^{-2}$ for ``peak $E_b$" and $\Phi_{t,b,\tau}=30^\circ$ and $90^\circ$; 
Higgs masses, couplings and widths computed with {\tt CPsuperH}.}
\end{figure*}

\begin{figure*}
\begin{tabular}{cc}
$\Phi_{t,b,\tau}=30^\circ$&
$\Phi_{t,b,\tau}=90^\circ$\\
\epsfig{file=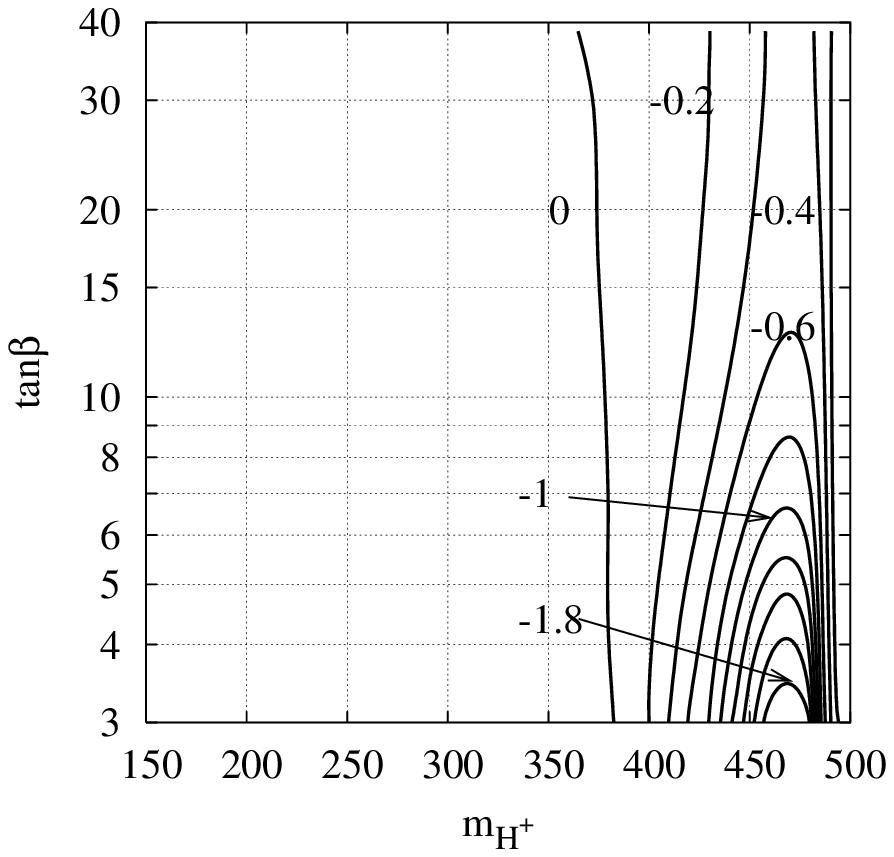,width=8.0cm}&
\epsfig{file=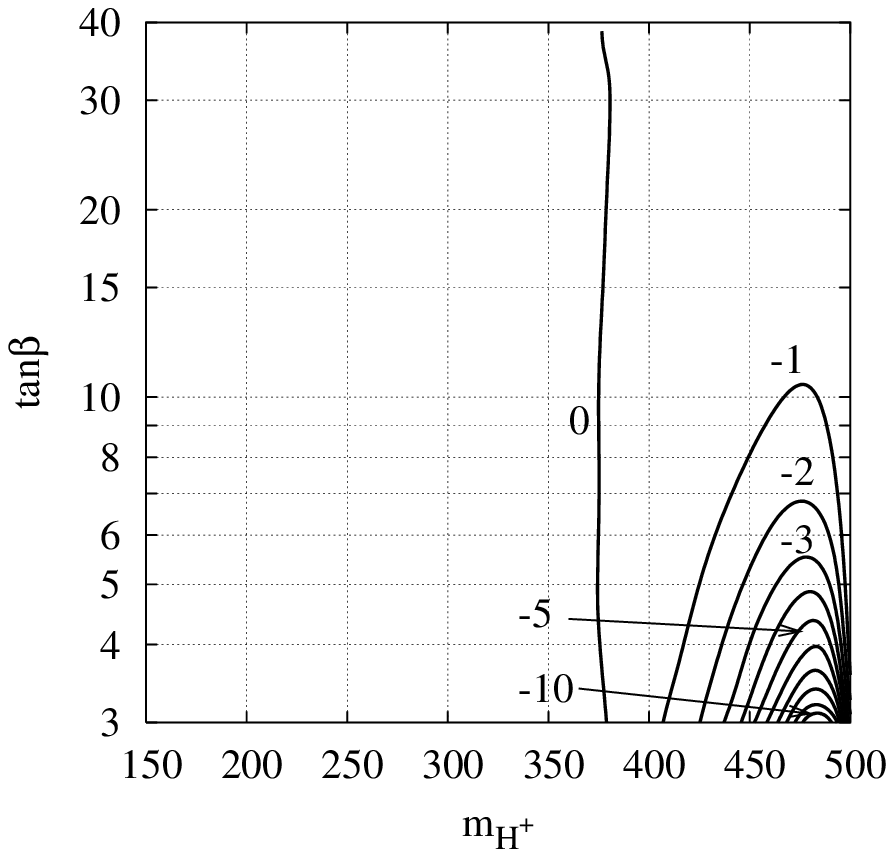,width=8.0cm} 
\end{tabular}
\caption{\label{fig:lept-fix-ch}Contours of constant ${\cal A}_3$ in units 
of $10^{-2}$ for fixed
$E_b=300$ GeV and $\Phi_{t,b,\tau}=30^\circ$ and $90^\circ$; 
Higgs masses, couplings and widths computed with {\tt CPsuperH}.}
\end{figure*}

The polarization of $\tau$ leptons can be measured using the energy distribution 
of the decay pions~\cite{Bullock:1991fd,Roy:1991sf,Bullock:1992yt,
Raychaudhuri:1995kv,Raychaudhuri:1995cc,rmg:stau}. The polarization of 
top quarks can be measured using energy distribution of 
$b$ quarks~\cite{Christova:1997by} 
or the angular distribution of decay leptons~\cite{Dalitz:1991wa, Arens:1992wh,
Espriu:2002wx,Godbole:2006tq}. 
This kind of analysis requires the full reconstruction of the
top momentum. Such a reconstruction may not always be possible for the 
semi-leptonic decay of the $t$ (or $\bar t$) quark. On the other hand, it is 
possible to construct simple asymmetries involving the polarization of the 
initial-state $e^\pm$ (and hence of the photons) and the charge of the 
final-state lepton, which are sensitive to CP violation. We denote the 
integrated cross section for the process 
$\gamma\gamma\to t\bar t \to l^+\nu b \bar t$ $(t l^-\bar\nu \bar b)$ by
$\sigma(\lambda_{e^-},Q_l)$, where $\lambda_{e^-}$ is the polarization 
of the electron beam in the parent collider and $Q_l$  the charge of the 
secondary lepton coming from the $t/ (\bar t)$ decay. The polarizations 
of all the other beams are adjusted to get a peaked spectrum and equal 
helicities for the incident photons. With this setup, we can define 
the following asymmetries~\cite{hel-God}:
\begin{eqnarray}
\mathcal{A}_1 = \frac{\sigma(+,+)-\sigma(-,-)}{\sigma(+,+)+\sigma(-,-)},
\nonumber\\
\mathcal{A}_2 = \frac{\sigma(+,-)-\sigma(-,+)}{\sigma(+,-)+\sigma(-,+)},
\nonumber\\
\mathcal{A}_3 = \frac{\sigma(+,+)-\sigma(-,+)}{\sigma(+,+)+\sigma(-,+)},
\nonumber\\
\mathcal{A}_4 = \frac{\sigma(+,-)-\sigma(-,-)}{\sigma(+,-)+\sigma(-,-)},
\label{lept-asym}
\end{eqnarray}
Only one of the above asymmetries is independent~\cite{hel-God} 
if  no cut is put on the lepton's polar angle in the laboratory frame. 
Even with a finite cut on the polar angle, the $\mathcal{A}_{1...4}$
have almost identical sensitivities to the Higgs couplings.
We  use a $20^\circ$ beam-pipe cut on the lepton.
The contours of constant ${\cal A}_3$ for $\Phi_{t,b,\tau}=30^\circ$ and  
$\ 90^\circ$, using {\tt CPsuperH}, are shown in Fig.~\ref{fig:lept-peak-ch} 
for the ``peak $E_b$" choice. 
Analogously, Fig.~\ref{fig:lept-fix-ch} shows ${\cal A}_3$ for fixed 
$E_b=300$~GeV and $\Phi_{t,b,\tau}=30^\circ$ and $90^\circ$.
The asymmetries  are sizable for $\Phi_{t,b,\tau}=90^\circ$ and decrease 
rapidly as $\Phi_{t,b,\tau}$ decreases.
For $\Phi_{t,b,\tau}=0^\circ$ the only source of CPV is the phase of $M_3$, 
$\Phi_3=90^\circ$, in our scenario.  All the $y_i$'s are then negligibly small 
as compared to the $x_i$'s, leading to very small values of the ${\cal A}_i$.
Here note that, as shown in Ref.~\cite{hel-God}, the lepton 
asymmetries of Eq.~(\ref{lept-asym}) are sensitive only to CP-odd combinations 
of the form factors, i.e.\ the $y_i$'s. 
This should be contrasted with the polarization observables, 
which are sensitive to both the CP-odd and CP-even combinations.


\section{Conclusions}
\label{six}

We have investigated the use of fermion polarization in the process 
$\gamma\gamma\to f\bar f$ (with $f=t$ or $\tau$) for studying neutral Higgs 
bosons at a photon collider. To this aim we have constructed polarization 
asymmetries $\delta P^\pm_f$, which are sensitive to the Higgs exchange 
contributions. We have also constructed a CP-odd asymmetry
$\delta P^{\rm CP}_f$ which is sensitive to CP violation in the Higgs sector.
All these asymmetries are constructed in a model-independent way and can be 
used to study Higgs bosons in various models beyond the SM.

We have applied this in a numerical analysis to the case of the MSSM with 
explicit CP violation. In particular we have evaluated our asymmetries 
for the CPX scenario, using the two public codes {\tt CPsuperH} and 
{\tt FeynHiggs} to calculate the Higgs masses, couplings and widths.

Scanning the $(\tan\beta-m_{H^+})$ plane for various phases $\Phi_{t,b,\tau}$, 
we found that $\delta P^\pm_\tau$ is sensitive to a light Higgs, 
especially if $\tan\beta$ is large. Assuming a measurement accuracy of 
$10^{-2}$  for $\tau$ polarization, $\delta P^\pm_\tau$ can in fact probe a 
large part of the 
CPV-MSSM parameter space. A cut on the $\tau\tau$ invariant mass is, however, 
necessary to enhance the signal. The CP-odd asymmetry $\delta P^{\rm CP}_\tau$, 
on the other hand, is always very small, well below measurability. 
While $\delta P^\pm_\tau$ can be enhanced by the cut mentioned above, 
this does not work for $\delta P^{\rm CP}_\tau$. 

This is complemented by the top polarization in $t\bar t$ production, which is sensitive to the heavier neutral Higgs bosons $\phi_{2,3}$, and also 
to CP mixing between them, for $m_{\phi_{2,3}}\ge 2m_t$ and small $\tan\beta$. 
A similar region is covered by the lepton asymmetries constructed from top decays, which are a pure measure of CP violation. These lepton asymmetries are large 
for large $\Phi_{t,b,\tau}$ but quickly decrease as this phase decreases.

To conclude, the top and tau polarization asymmetries presented in this paper 
may prove useful to study the effects of a CP-violating Higgs sector 
at a photon collider. They may in particular cover the parameter region where 
a light CP-violating Higgs may have been missed at LEP, and may be missed as 
well at LHC.
We found, however, large quantitative differences between the results
obtained with {\tt CPsuperH} and {\tt FeynHiggs}. In this regard we emphasize 
the need for a standardization of these tools.

\section*{Note added}While this paper was in 
preparation, at a point where the numerical analysis was already finished, 
a new version of {\tt FeynHiggs} was released, see
{\tt http://www.feynhiggs.de} and  the contribution on {\tt FeynHiggs} in
\cite{cpnsh}.
This version contains new radiative corrections also for the CP-violating case. 
It will be interesting to see their effect on the polarization observables 
discussed in this paper.

\section*{Acknowledgments}

We thank S. Heinemeyer for discussions regarding {\tt FeynHiggs}. 
R.M.G., S.D.R.\ and R.K.S.\ acknowledge the support of the Department of 
Science and Technology, India, under project no. SP/S2/K-01/2000-II
and of the Indo-French Center for the Promotion of Advanced Research 
under IFCPAR project no. IFC/3004-B/2004.
The work of S.K.\ is financed by an APART (Austrian Program for Advanced 
Research and Technology) grant of the Austrian Academy of Sciences.


\end{document}